\documentclass[aps,prb,twocolumn,groupedaddress,amsmath,amssymb,showkeys]{revtex4}

\usepackage{graphicx}
\usepackage{hyperref}
\usepackage{pifont}
\usepackage{color}

\begin{document}

\title{Electric Field Effects on Armchair MoS$_2$ Nanoribbons}

\author{Kapildeb Dolui, Chaitanya Das Pemmaraju and Stefano Sanvito}
\affiliation{School of Physics and CRANN, Trinity College, Dublin 2, Ireland}

\date{\today}

\begin{abstract}
\centerline{\bf Abstract}
{\it Ab initio} density functional theory calculations are performed to investigate the electronic structure of MoS$_2$ armchair 
nanoribbons in the presence of an external static electric field.  Such nanoribbons, which are nonmagnetic and semiconducting, 
exhibit a set of weakly interacting edge states whose energy position determines the band-gap of the system. We show that, by 
applying an external transverse electric field, $E_\mathrm{ext}$, the nanoribbons band-gap can be significantly reduced, leading 
to a metal-insulator transition beyond a certain critical value. Moreover, the presence of a sufficiently high density of states at the 
Fermi level in the vicinity of the metal-insulator transition leads to the onset of Stoner ferromagnetism that can be modulated, and 
even extinguished, by $E_\mathrm{ext}$. In the case of bi-layer nanoribbons we further show that the band-gap can be changed 
from indirect to direct by applying a transverse field, an effect which might be of significance for opto-electronics applications. 
\end{abstract}


\maketitle

Over the past several years one dimensional (1D) nanostructures, such as nanotubes, wires, rods, belts and ribbons have attracted 
a growing interest from researchers keen to investigate the wide array of photophysical, photochemical and electron-transport properties 
that are unique to their dimensionality.\cite{review_1D_2003, ACS_Ich_2006_45} The study of these nanostructures has also been facilitated 
by recent advances in nano-lithographic techniques,\cite{review_AdvP_3} such as electron beam or focused-ion-beam (FIB) 
writing~\cite{review_AdvP_4a,review_AdvP_4b} and X-ray or extreme-UV lithography,\cite{review_AdvP_5} whereby such 1D systems 
can be readily fabricated in the research laboratory. From a nanotechnology perspective, 1D structures offer a range of potential applications 
that are different from those provided by their 2D and 3D counterparts.\cite{review_1D_2003,review_AdvP_1}

While carbon nanotubes (CNTs) remain the most widely studied 1D nanostructures to date, nanowires (NWs) and nanoribbons (NRs) have 
lately received increasing attention as possible alternatives. In particular the fact that the electronic structure of NRs can be modified by 
manipulating their edges, which usually are more reactive than the bulk, offers a powerful tool for customizing such nanostructures to a 
particular application. For this reason recent times have witnessed an explosion of theoretical and experimental studies on NRs. Primarily these 
have been devoted to graphene NRs (GNRs),\cite{PRB_54_17954,PRB_59_8271,PRL_2007_98}
but many other materials have been either made or predicted in the NR form. These include 
BC$_3$,\cite{APL_2009_94,JACS_133_16285}
BN,\cite{NL_2008_8, NL_2008_8_2,NL2011113221} 
ZnO,\cite{Pan_2001,ChPL_2007_448,ACS_Nano_2010_4} 
Si,\cite{JACS_123_11095,APL_2009_95} {\it etc}. 
Intriguingly for some of these a magnetic~\cite{ACS_Nano_2010_4,APL_2009_95} or even a half-metallic ground state has been 
predicted.\cite{PRB_2008_78}

This work investigates MoS$_{2}$ NRs, which represent one of the several low-dimensional structures that can be made from 
transition-metal dichalcogenides. Layered transition-metal dichalcogenides are particularly interesting because of the large variety of 
electronic phases that they can exhibit,\cite{ReviewMoS2-1,ReviewMoS2-2} namely metallic, semiconductor, superconductor and 
charge density wave. Bulk MoS$_2$ has a prototypical layered structure where Mo is covalently bonded to S with a trigonal prismatic 
coordination. Each S-Mo-S sandwich layer is tightly bound internally and interacts weakly with the neighboring sandwich only through 
van der Waals forces.\cite{SSCom_1972_11} Because of such a structure the fabrication of ultrathin crystals of MoS$_2$ is possible by micro-
mechanical cleavage~\cite{PNAS_2005_102} or exfoliation.\cite{JNC_2011} Therefore, like graphene,\cite{Sc_2004_306} single layers of 
MoS$_2$ can be extracted repeatedly one by one from bulk materials and deposited on substrates for further studies.\cite{arxive_1}

From the parental MoS$_2$ single-layer crystal several nanostructures can be made. These have been traditionally studied in the context of 
catalysis for desulfurization processes~\cite{NatNanoT_2007_2, Jcat_2007_249} and as thermoelectric materials.\cite{JPChemC_2007_111} 
Here we explore a different aspect, namely how the electronic properties of MoS$_2$ armchair nanoribbons (ANRs) can be manipulated by 
the application of an external electric field, $E_\mathrm{ext}$. In particular we look at the possibility of inducing a metal-insulator transition in 
the nanoribbons, and at the associated magnetic moment formation via the Stoner mechanism in the search for a large magneto-electric effect. 
Our study thus complements those already reported in the literature for graphene,\cite{nature_2006_444} BN,\cite{NL_2008_8}
BC$_3$~\cite{APL_2009_94} and AlN.\cite{CPL_2009_469}

The paper is organized as follows. In the next section we present a description of the various structures investigated. The calculated 
electronic properties of bulk and single layer MoS$_2$ as well as several ANRs are discussed in the following section. First we analyze 
single layer MoS$_2$ ANRs and present results from non spin-polarized calculations including an applied static electric field. These are 
explained by means of a simple tight-binding model. Then the electronic structure and the electric field response of bilayer and multi-layer 
ANRs are presented. Finally we show results obtained from spin-polarized calculations investigating magneto-electric effects in single 
layer MoS$_2$-ANRs and, before concluding, we consider the effects that different edge terminations have on the electric field driven 
magnetism.

\section*{NANORIBBONS STRUCTURE}\label{Comp}

Bulk MoS$_2$ has a hexagonal crystal structure with space group P6$_3$/mnc (D$_{6h}^4$) and it is the 2D template for constructing 
the NRs. Similarly to C nanotubes,\cite{nature_injma} MoS$_2$ NRs may be described by the 2D primitive lattice vectors $\vec{a}$ 
and $\vec{b}$ of the parental 2D structure and two integer indices $(n,m)$,\cite{PRL_68_1597} so that the chiral vector is defined 
as $\vec{C_h} = n\vec{a} + m\vec{b}$. Three types of NRs can thus be identified: {\it zig-zag} for $n = m$, 
{\it armchair} for $n\neq 0$, $m = 0$, and {\it chiral} for $n\neq m$. MoS$_2$ ANRs are nonmagnetic semiconductors irrespective of their size, 
whereas the zig-zag nanoribbons (ZNRs) are predicted metallic and magnetic.\cite{JACS_2008_130} Since our goal is that of describing an 
electric-field induced metal-insulator transition our starting point must consist in NRs with an insulating ground state. As such we consider only 
ANRs. As a matter of notation, following several previous studies,\cite{JACS_2008_130,NL_2008_8,NLStefano} we identify the different 
sized MoS$_2$ ANRs as $n$-ANR, where $n$ is the number of dimer lines across the terminated direction of the 2D MoS$_2$ layer, {\it i.e.} 
across the non-periodic dimension of the nanoribbon~(see Fig. \ref{fig:10-ANR.eps}). The multilayer ribbons are constructed by 
placing single layer ribbons on top of each other with an ABA stacking.\cite{PRB_2009_79}

\section*{RESULTS AND DISCUSSION}
\subsection*{Electronic properties}\label{ElecP}
Our systematic study begins with calculating the electronic properties of MoS$_2$ in its bulk form. The optimized bulk MoS$_2$ unit cell 
parameters are $a=b=3.137$~{\AA}, $c/a=3.74$, while the S-Mo-S bonding angle is 82.64$^{\circ}$. These values are in good agreement 
with previous theoretical calculations~\cite{PRB_2001_64_235305} and also with the experimental ones of $a=b=3.16$~{\AA}, 
$c/a=3.89$.\cite{JPCM_1992_4} The Mo-S bond length in bulk MoS$_2$ is found to be 2.42~{\AA},  again in close agreement with the 
experimental value of 2.41~{\AA} \cite{JPCM_1992_4} and to the earlier theoretical estimate of 2.42~{\AA}.\cite{ReviewMoS2-1} 
Bulk MoS$_2$ is a semiconductor and we predict an indirect band gap of 0.64~eV between the $\Gamma$ point and a point half way 
along the $\Gamma$-K line. Our calculated band-gap is smaller than the experimental one of 1.23~eV, \cite{PRB_2001_64_235305,JPCM_1992_4}
but it is in good agreement with existing density functional theory (DFT) calculations at the local spin density approximation (LSDA) 
level.\cite{PRB_2001_64_235305,PRB_1995_51} It is well known that the LSDA systematically underestimates the band-gap, so that such 
result is not surprising. However, we note here that the LSDA underestimation affects our results only at a marginal quantitative level.
\begin{figure}
\center
\includegraphics[width=8.0cm,clip=true]{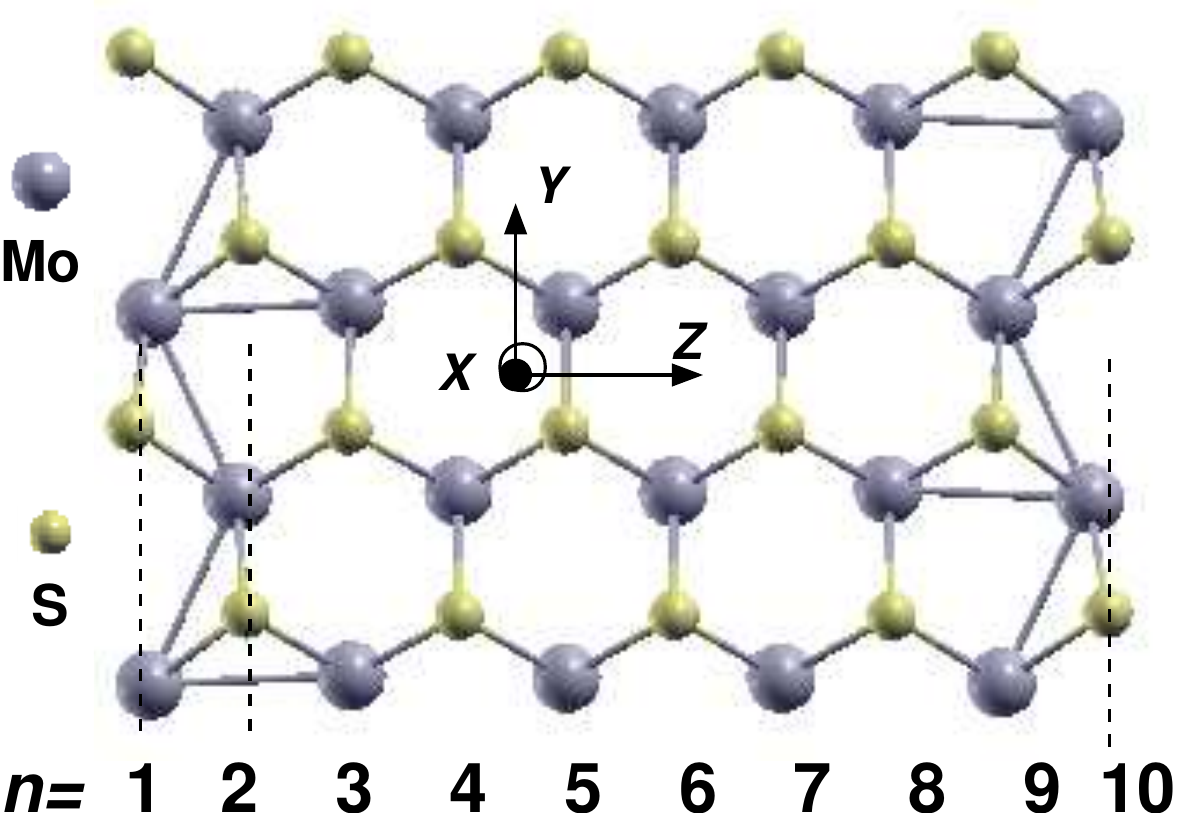}
\caption{(Color online) The optimized structure of a MoS$_2$ 10-ANR. The ribbon is periodic along the $y$ direction. Color code: 
grey (dark grey) = Mo, yellow (light grey) = S.}
\label{fig:10-ANR.eps}
\end{figure}

Next we move to study the electronic properties of a single MoS$_2$ layer. Our optimized lattice constant, $a=b$, is now 3.132~{\AA}, {\it i.e.} it is 
slightly smaller than that in the bulk. Such value is in close agreement with the experimentally observed one of 3.15~{\AA}.\cite{PRL_2000_84} 
Our calculations show that as the number of layers is decreased from the bulk to a few layers, the minimum of the lowest unoccupied band shifts 
from half way along the $\Gamma$-K line to K, with a single MoS$_2$ layer exhibiting a direct band-gap at K. In this context, recent 
experiments~\cite{NL_2010_10,PRL_2010_136805} have shown that as the thickness of layered MoS$_2$ samples decreases from the bulk 
towards the monolayer limit, photoluminescence emerges, indicating the transition from an indirect to a direct band-gap. A similar conclusion 
was reached by comparing scanning photoelectron microscopy to DFT calculations.\cite{PRB_2011_045409} Both for the bulk and the single 
layer the band structure around the Fermi level, $E_\mathrm{F}$, is derived mainly from Mo-4$d$ orbitals, although there are smaller contributions 
from the S-3$p$ via hybridization within the layer.

In Fig.~\ref{fig:10-ANR.eps} the optimized geometry of a MoS$_2$ 10-ANR is shown (the periodicity is along the $y$-direction). 
For symmetric ANRs ($n$ odd) the two edges have mirror refection symmetry, while this is not the case for the antisymmetric ones ($n$ even). 
Our calculations show that the two possible ribbon configurations are essentially energetically degenerate, meaning that the total energy per 
atom scales with the ribbon size but does not depend on the ribbon symmetry. We have then checked that the electronic properties and their 
dependence on the external electric field are insensitive to the ribbon geometry and here we present data only for the asymmetric case.
\begin{figure}
\centering
\includegraphics[width=8.0cm,clip=true]{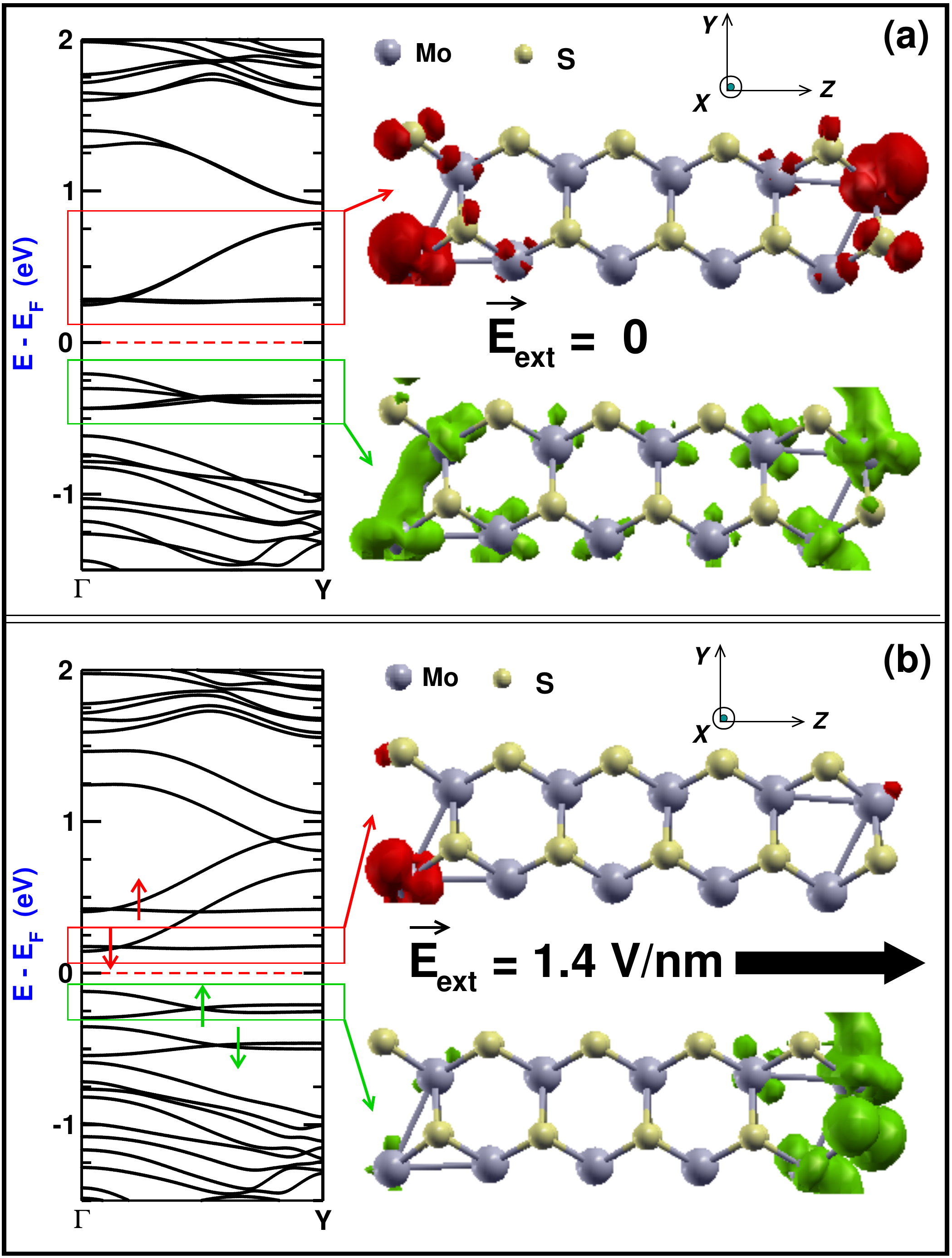}
\caption{(Color online) Electronic structure of a 10-ANR as a function of an external transverse electric field, $E_\mathrm{ext}$. In (a) we 
present: (left) the band structure and (right) the LDOS respectively of the CB (top) and VB (bottom) for $E_\mathrm{ext}=0$. In (b) the same 
quantities are shown for $E_\mathrm{ext} = 1.4$~V/nm. The LDOS are taken over the energy range indicated by the red (dark gray) and 
green (light gray) boxes respectively above and below the Fermi level (see band-structure). Note that in both cases the CB and VB are characterized 
by states located at the NR edges. }
\label{fig:band+LDOS}
\end{figure}

In order to determine the ground state of the different $n$-ANRs, we have first carried out both spin-unpolarized and spin-polarized total 
energy calculations including geometry optimization. We take the case of a 10-ANR as an example and we use its electronic structure to 
discuss the general properties of such a NRs class. From the bandstructure [shown in Fig.~\ref{fig:band+LDOS}(a)], it is clear that a 10-ANR 
is a non-magnetic semiconductor with a direct (LSDA) band-gap of 0.45~eV at the $\Gamma$ point. This is in agreement with previous 
calculations.\cite{JACS_2008_130} The local density of states (LDOS) of the conduction band (CB) and valence band (VB) [shown in 
Fig.~\ref{fig:band+LDOS}(a)] indicates that the electronic states around $E_\mathrm{F}$ are completely localized at the ANR's edges. 
The CB and the VB originate from a hybridized mix of Mo-4$d$ and S-3$p$ orbitals with the hybridization being stronger in the VB than 
in the CB. 

Such results are relatively independent of the ribbon size, and unlike graphene NRs, all the MoS$_2$ ANRs are semiconducting. For the 
smaller $n$-ANRs ($n\leq24$), the band-gap oscillates in magnitude with increasing $n$ and finally converges to a constant value of 
around 0.52~eV for larger sized ribbons ($n>24$). The same oscillatory behaviour has been observed in earlier calculations 
\cite{JACS_2008_130} and it is quite similar to that predicted for BN-ANRs.\cite{NL_2008_8_2} As $n$ increases we also observe 
oscillations in the equilibrium lattice constant, which slowly approaches a constant value of $\sim$ 3.132 {\AA}, similar to that calculated for 
the infinite MoS$_2$ single layer. We thus note that the calculated band-gaps for all $n$-ANRs are much smaller than that of the infinite 
MoS$_2$ single layer~(1.90 eV), while the lattice constants deviates only marginally. The reason for such a difference is rooted in the fact that 
both the VB and CB of the ribbons are formed by states strongly localized at the two edges. Indeed these states do not exist in the case of the 
infinite MoS$_2$ single layer and they are simply a consequence of the different wave-function boundary conditions. 
\begin{figure}
\centering
\includegraphics[width=8.0cm,clip=true]{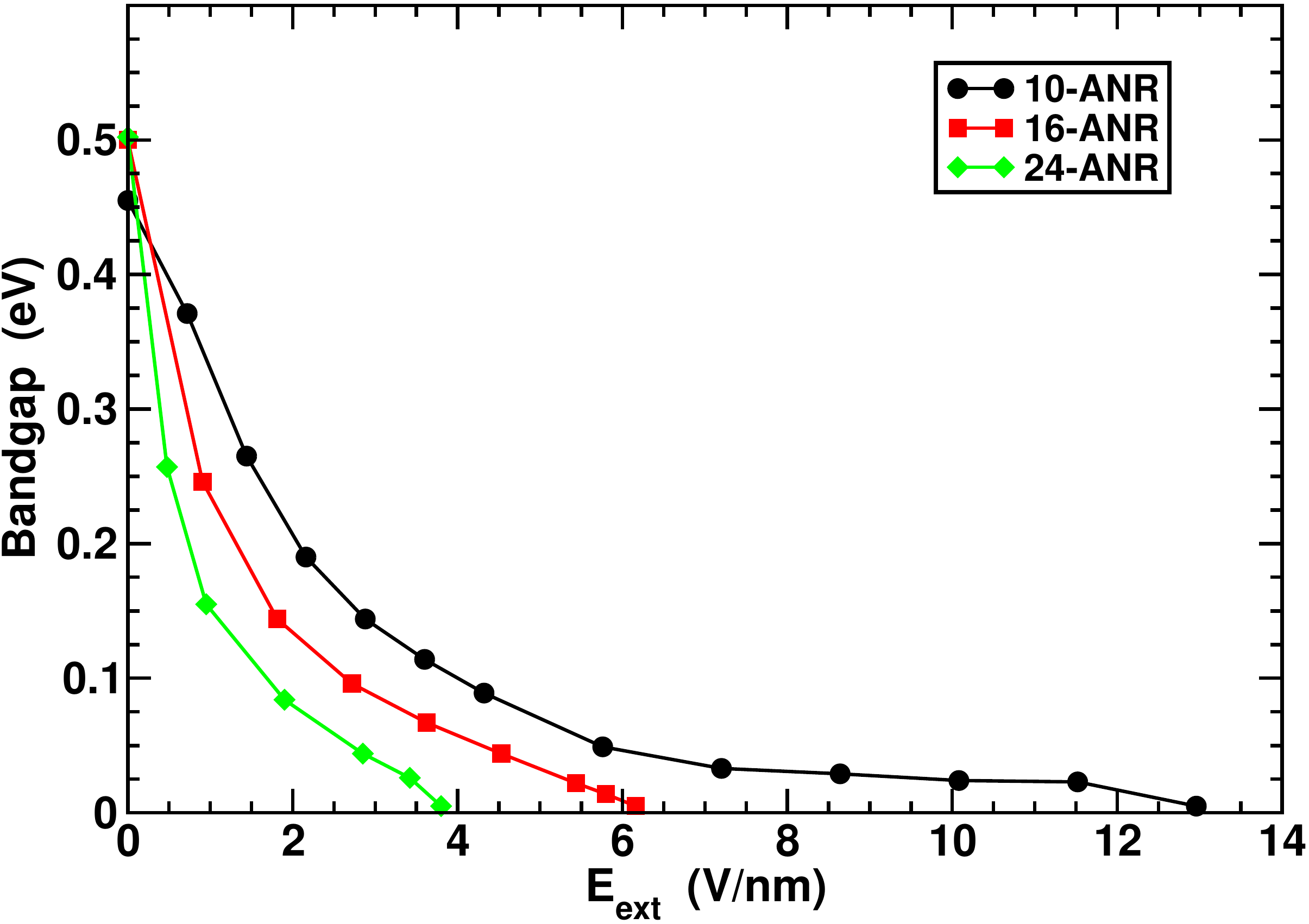}
\caption{(Color online) Variation of the elementary LSDA band-gap with the applied transverse electric field, $E_\mathrm{ext}$, for 
10-ANR (black circles), 16-ANR (red squares), 24-ANR (green diamonds).}
\label{fig:BGvsE_NR}
\end{figure}

\subsection*{Response of a MoS$_2$-ANR to $E_\mathrm{ext}$}\label{Resp}
We now discuss the response of the electronic structure of the MoS$_2$-ANRs to a static external electric field, $E_\mathrm{ext}$. 
As mentioned earlier, the size of the ribbon band-gap is determined by the energy position of the edge states forming the CB and the VB. 
Thus any change in the ANR band-gap under an applied field would be mainly determined by the response of its edge states. It is worth 
noting that for $E_\mathrm{ext}=0$, both the CB and VB are doubly degenerate as there are two states in each band corresponding to the 
two edges of the ribbon (this means that the electron density corresponding to either the CB or the VB is equally distributed over the two 
opposite edges). We find no gap modulation when $E_\mathrm{ext}$ is perpendicular to the the plane of ribbon, indicating that a planar 
MoS$_2$ nano-structure with a longitudinal gate will not be electronically responsive. In contrast a significant modulation of the band-gap 
can be obtained by means of a transverse field. This is applied along the $z$-direction according to the geometry of Fig.~\ref{fig:10-ANR.eps}. 
In practice in our calculations a periodic sawtooth-type potential perpendicular to the ribbon edge is used to simulate the transverse electric 
field in the supercell so that the potential remains homogeneous along the ribbon edges.\cite{PRB_1988_38}

As the transverse $E_\mathrm{ext}$ is applied the band-gap decreases monotonically while remaining direct at $\Gamma$ 
[see Fig.~\ref{fig:band+LDOS}(b) and Fig.~\ref{fig:BGvsE_NR}]. Such a behaviour can be understood by assuming little interaction between 
the electron densities at the two edges. Under this assumption the only effect produced by a transverse electric field is that of creating an 
electrostatic potential difference across the ribbon. As a consequence, the band manifold (either belonging to the VB or the CB) localized at 
the edge kept at the higher external potential moves upwards in energy, while that kept at the lower potential moves in the opposite direction. 
Hence, the new band-gap of the system in the presence of an external electric field is formed between the CB manifold localized at the lower 
potential edge and the VB manifold localized at the opposite one. The edge degeneracy is thus broken. As the field strength increases the 
band-gap reduces further and eventually vanishes for a critical field, $E_\mathrm{c}$, characteristic of the specific nano-ribbon. 
Note that such a Stark-driven gap modulation has been previously reported for C nanotubes \cite{APL_2002_80} and for nano-ribbons made 
of different materials such as graphene,\cite{nature_2006_444} BN,\cite{NL_2008_8, NL_2008_8_2,PRB_2008_77} AlN,\cite{CPL_2009_469} 
{\it etc}. Notably, in the case of MoS$_2$ and in contrast to some other compounds such as BN,\cite{PRB_2008_77} the gap closure is independent 
from the field polarity, reflecting the perfect mirror symmetry of the ribbon's edges.  

\begin{figure}
\centering
\includegraphics[width=8.0cm,clip=true]{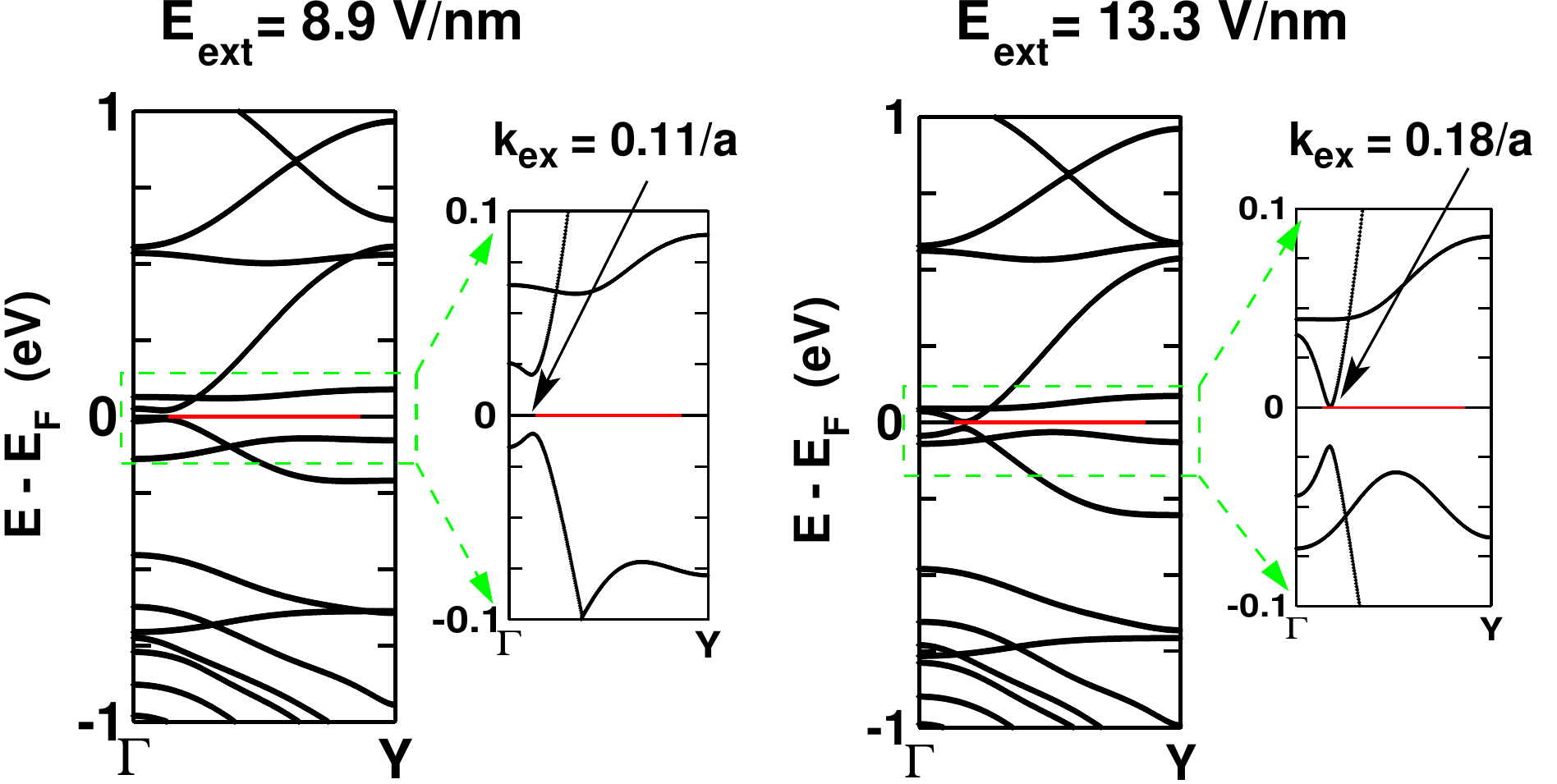}
\caption{(Color online) Non-spin polarized band-structure of a MoS$_2$ 10-ANR in presence of a transverse field of magnitude close to that 
needed for the gap closure: $E_\mathrm{ext}=8.9$~V/nm (left) and $E_\mathrm{ext}=13.3$~V/nm (right). The smaller figures are a zoom 
around the Fermi level (a is the lattice constant).}
\label{fig:Band_NR5_E_6to9}
\end{figure}
Figure~\ref{fig:BGvsE_NR} shows the evolution of the band-gap as a function of the external electric field for three selected nano-ribbons, 
respectively 10-ANR, 16-ANR and 24-ANR. In general we observe that the band-gap drops more rapidly with $E_\mathrm{ext}$ as the size of 
the $n$-ANRs gets larger. Such a width dependence can be easily rationalized by assuming again little interaction between the two ribbon 
edges. In this case the potential difference between the edges necessary to close the gap is the same regardless of the ribbon size. If one now 
assumes that the potential drop inside the ribbon is approximately uniform (linear), we will conclude that larger ribbons necessitate smaller 
electric fields to sustain the same potential difference at the edges. As such the critical field, $E_\mathrm{c}$, decreases with the ribbon width 
and already for a 24-ANR it assumes a value around 4~V/nm. 

Interestingly our calculated values for the critical field $E_\mathrm{c}$ are quite similar to those obtained before for BN~\cite{PRB_2008_77} and 
AlN,\cite{CPL_2009_469} despite the fact that the band-gaps in these materials are much larger. Such a fact however should not be surprising.           
In fact, the band-gap closure occurs because of the almost rigid shift of the edge-localized ribbon CB and VB when the field is applied. 
As such, the condition for gap closure is that the external field produces a potential difference, $\Delta V$, at the nano-ribbon edges that 
matches the ribbon band-gap, $\Delta E_\mathrm{g}$, {\it i.e.} $e\Delta V=\Delta E_\mathrm{g}$, where $e$ is the electron charge. Under the 
assumption of a linear potential drop (constant electric field) inside the ribbon, we obtain the relation $e\Delta V=eE_\mathrm{ext}\frac{d}{\kappa}$, 
where $\kappa$ is the ribbon dielectric constant along the transverse direction and $d$ is the ribbon width. The critical field for the gap closure 
then simply reads
\begin{equation}\label{Eq1}
\Delta E_\mathrm{g}=e\Delta V=E_\mathrm{c}\frac{ed}{\kappa}\;\;\;\;\rightarrow\;\;\;\;E_\mathrm{c}\propto\frac{1}{ed}\;,
\end{equation}
where the second equality follows from the fact that the dielectric constant is approximately inversely proportional to the material band-gap. 
The equation (\ref{Eq1}) brings two important consequences. On the one hand, it tells us that the critical field for the gap closure is 
approximately materials independent. On the other hand, it establishes a $1/d$ decay of $E_\mathrm{c}$ with the ribbon width.

Before the band-gap closes completely with increasing $E_\mathrm{ext}$ an interesting effect is observed in small sized ANRs 
(for example in 10-ANR), namely that the gap remains direct but it moves away from $\Gamma$ towards Y in the 1D Brillouin zone. This 
shift occurs simultaneously with the band-gap reduction and it is seen to become more pronounced as $E_\mathrm{ext}$ gets larger. 
Such an effect can be observed in Fig.~\ref{fig:Band_NR5_E_6to9} where the band-structure of the 10-ANR is plotted for two different 
values of the electric field. The band-gap shift away from $\Gamma$ appears because of the interaction between the two edges of the 
ribbon and can be explained with the help of a simple tight-binding model, which we develop next.

As already mentioned before, the CB and VB are extremely localized at the edges of the ANR, so that their dispersion is solely 
determined by the longitudinal dimension. Then we can model their electronic structure by considering a simple nearest-neighbour tight binding 
model for two linear chains (mimicking the two 1D edges). For the sake of simplicity, we take only $s$ orbitals in the model and 
the CB and VB are simply characterized by two different on-site energies, respectively $\epsilon_\mathrm{L}$ and $\epsilon_\mathrm{R}$ 
(L and R stand for left and right-hand side edge). This is of course a rather crude model, as both $p$ and $d$ orbitals are excluded. 
However, the band-gap closure and the formation of a magnetic moment both originate from the one-dimensional nature of the edge 
states and not from the details of their orbital composition. As such our simple model captures the essential features of this problem,
while further details related to edge specific nature will be discussed later in the paper.
\begin{figure}[htbp]
\centering
\includegraphics[width=8.5cm,clip=true]{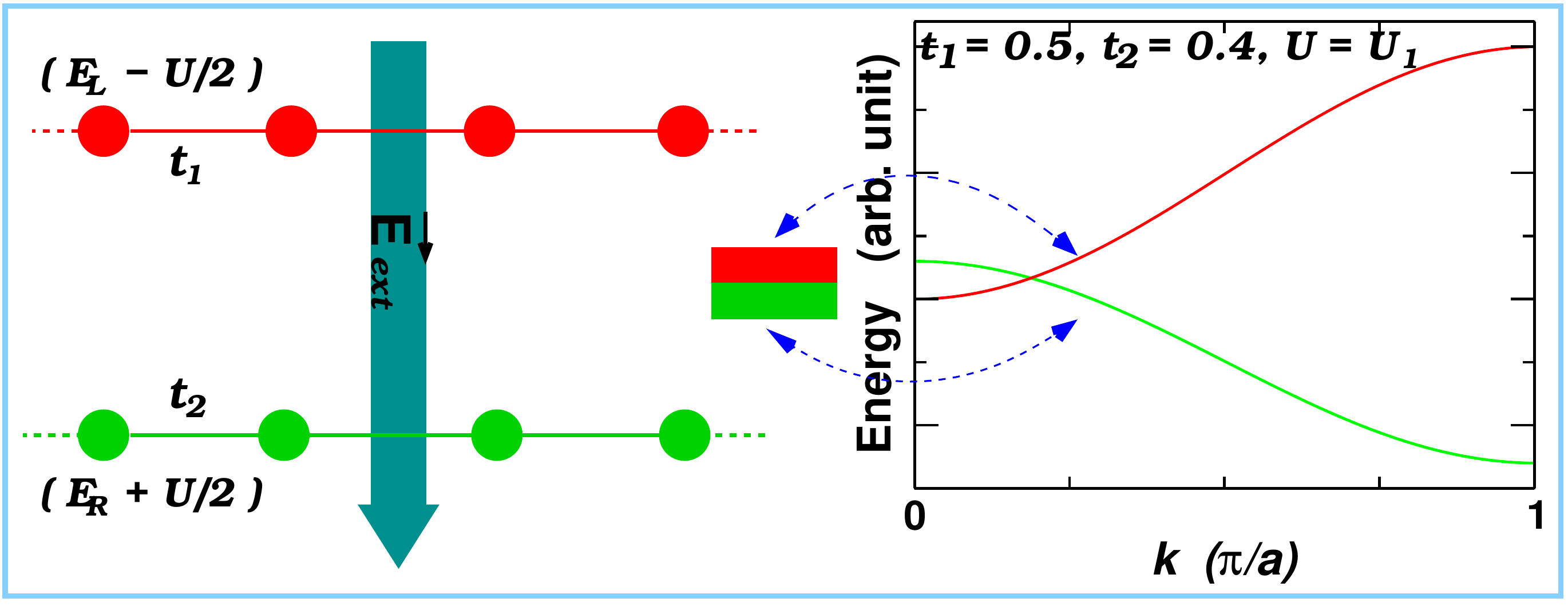}
\includegraphics[width=8.5cm,clip=true]{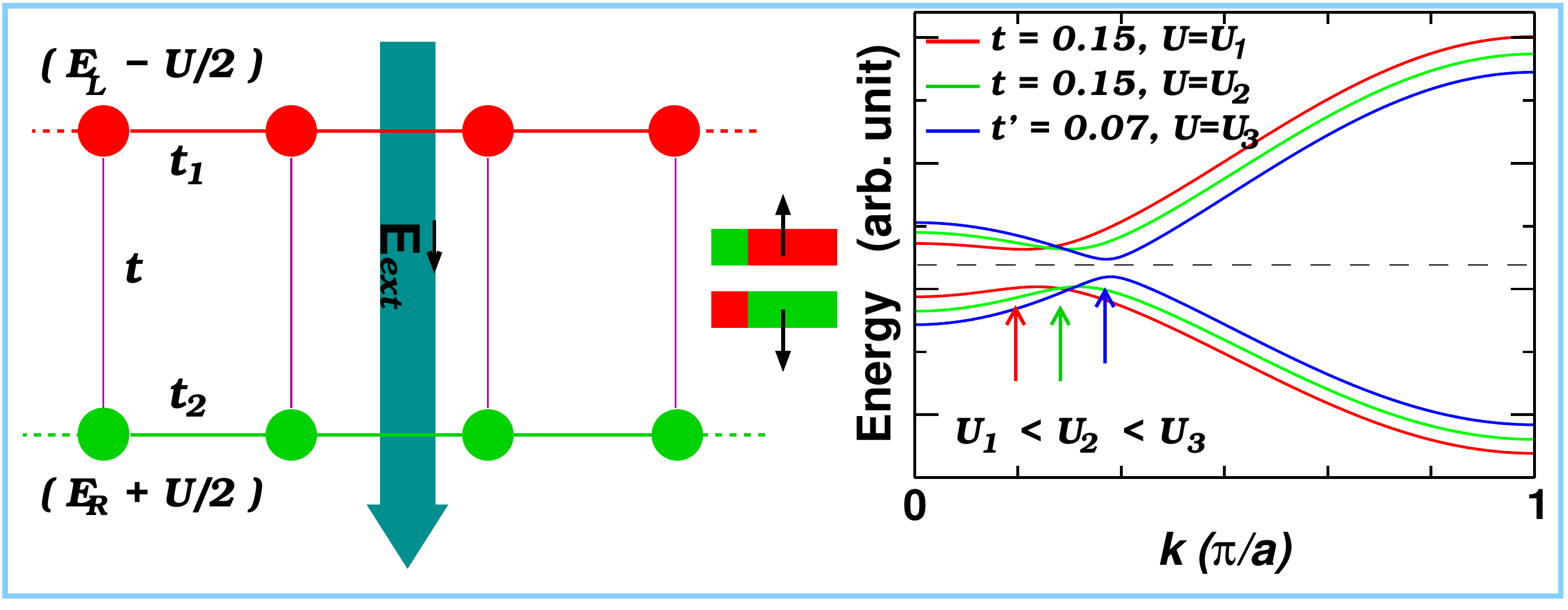}
\caption{(Color online) The band-structure of two linear chains calculated by using a simple tight-binding model. In the top panel the interaction
between the chains is assumed to vanish, while in the bottom one there is an additional hopping matrix element between atoms belonging to different
chains (the chains are arranged on a square lattice). The horizontal dotted line in the lower panel corresponds to the Fermi level.}
\label{fig:TB_2}
\end{figure} 

The two edges of a nanoribbon interact with each other in two possible ways. On the one hand, electrons can tunnel between the two edges with a 
probability given by the hopping integral $t$. This is expected to decrease as the ribbon width becomes larger. On the other hand,
upon the application of an external field, the bond-charges at the two ribbon edges will interact electrostatically. Such an interaction
is taken into account by the dielectric response of the ribbon, which is described by the transverse component of the dielectric
constant $\kappa$. As such, the potential difference between the ribbon edges, $\Delta V$, is related to the external field simply
as $\Delta V=E_\mathrm{ext}\frac{d}{\kappa}$.

The 1D band-structures for the two edges are simply $E_\mathrm{L}^k=\epsilon_\mathrm{L}-2t_1\cos k$ and 
$E_\mathrm{R}^k=\epsilon_\mathrm{R}+2t_2\cos k$, where $t_1>0$ and $t_2>0$ are the hopping integrals respectively of the left- and 
right-hand side chain, and $k$ is the 1D adimensional wave-vector (see Fig.~\ref{fig:TB_2}). Let us assume that 
$\epsilon_\mathrm{L}>\epsilon_\mathrm{R}$ so that the left-hand side edge corresponds to the CB and the right-hand side one to the 
VB (the band-gap is at $\Gamma$). Let us also assume for the moment that there is no inter-chain interaction, {\it i.e.} that the hopping
integral between the two chains vanishes, $t=0$. Clearly, if $|\epsilon_\mathrm{L}-\epsilon_\mathrm{R}|>2(t_1+t_2)$ there will be a gap 
between the CB and VB. The presence of an electric field simply shifts the on-site energy of the two bands. Thus the new on-site energies 
will be respectively $\epsilon_\mathrm{L}-U/2$ and $\epsilon_\mathrm{R}+U/2$, with $U=e\Delta V$. This simple model then predicts that 
the band-gap will close for $U=\epsilon_\mathrm{L}-\epsilon_\mathrm{R}-2(t_1+t_2)$. For electric fields exceeding such value the ribbon 
will appear as a semi-metal, {\it i.e.} it will present coexisting electron and hole pockets at the $\Gamma$ point. 

Let us now investigate the situation in which there is inter-chain interaction, {\it i.e.} $t\ne0$ between atoms localized on different chains (the 
atoms are assumed to be arranged on a square lattice). The new band-structure now takes the form
\begin{equation} \label{Ek}
{E^k_\pm}=\frac{E_\mathrm{L}^k + E_\mathrm{R}^k}{2} \pm \frac{1}{2} \sqrt{[E_\mathrm{L}^k - E_\mathrm{R}^k - U]^2 + 4t^2}\:,
\end{equation} 
where the ``$+$'' sign is for the CB and the ``$-$'' one is for the VB. Clearly inter-chain hopping opens up a band-gap (of size 2$t$) at the point 
along the $\Gamma$-Y line, where the two bands would otherwise cross for $t=0$ (as shown in Fig. \ref{fig:TB_2}). It also indicates that, if the 
applied electric field increases further ($U$ gets larger), the $k$-point where the direct band-gap appears will shift towards Y, but the value of the 
band-gap itself will remain constant. It then follows that the band-gap closure occurs only if $t$ is reduced simultaneously as $E_\mathrm{ext}$ 
is increased (see the curve in Fig.~\ref{fig:TB_2} for $t'< t$). This essentially suggests that the polarization of the edge state wave-functions under 
the influence of $E_\mathrm{ext}$ occurs in such a way as to reduce the effective interaction between the two edges of the ribbon. Notably the 
position in $k$-space of the band-gap, $k_\mathrm{ex}$, can be found by minimizing Eq.~(\ref{Ek}). This gives us
$k_\mathrm{ex}=\cos^{-1}\left[\frac{U_0-U}{2(t_1+t_2)}\right]$, with $U_0=\epsilon_\mathrm{L}-\epsilon_\mathrm{R}$. 
\begin{figure}
\centering
\includegraphics[width=8.0cm,clip=true]{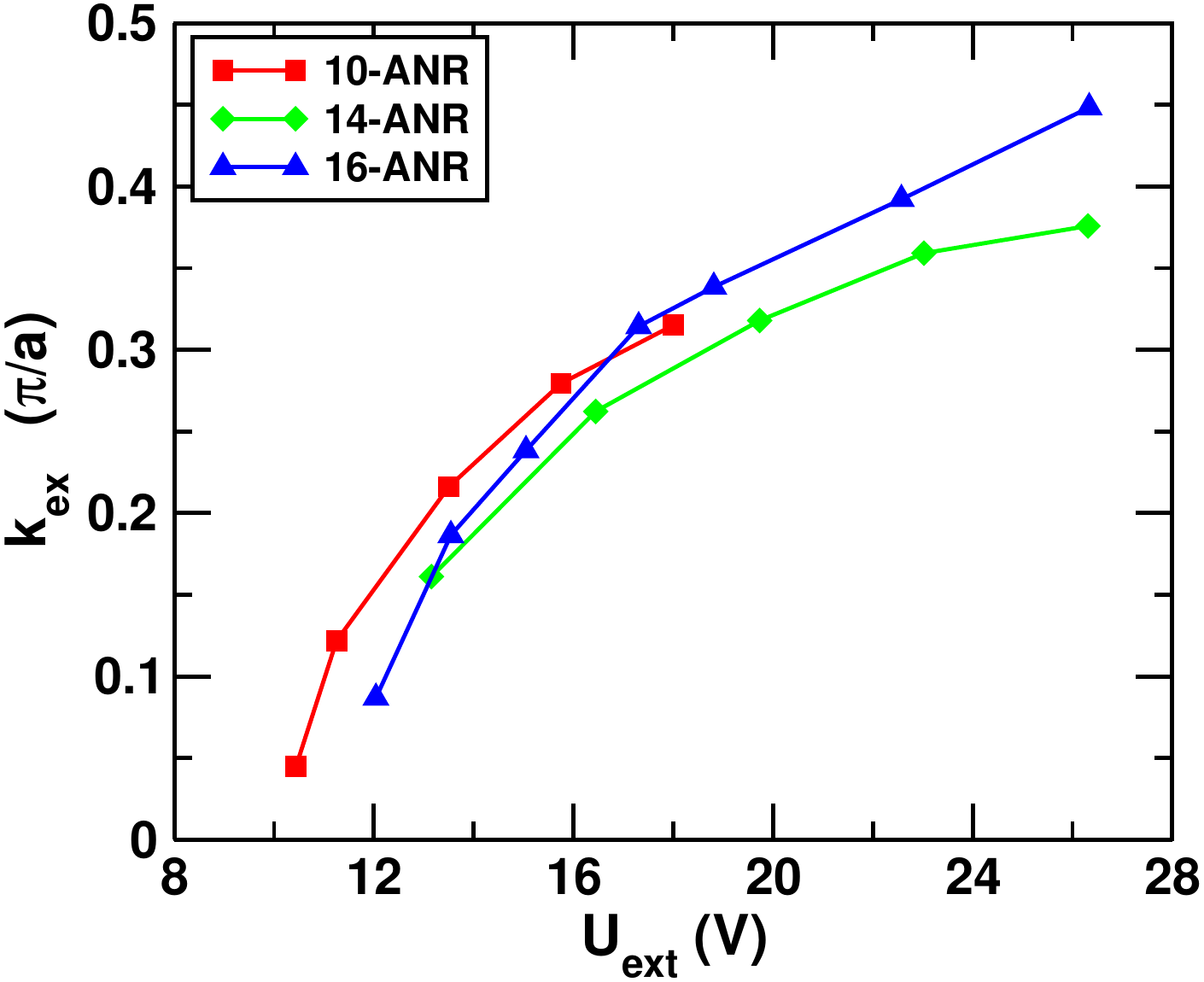}
\caption{(Color online) The variation of the $k$-vector, $k_\mathrm{ex}$, corresponding to position of the energy band-gap as a function of 
applied bias for different $n$-ANRs. Here $U_\mathrm{ext}=E_\mathrm{ext}d$, where $d$ is the nano-ribbon width.}
\label{fig:Kc_U}
\end{figure}
Such a qualitative picture agrees quite well with our DFT calculated $k_\mathrm{ex}$, which is presented in Fig. \ref{fig:Kc_U} for small 
ANRs ($n < 10$). From the figure it is also worth noting, again in agreement with our simple model, that $k_\mathrm{ex}$ is practically 
independent from the ribbon size once the curve is plotted as a function of the external potential at the ribbon edges 
$U_\mathrm{ext}=E_\mathrm{ext}d$.

\begin{figure}
\centering
\includegraphics[width=7.0cm,clip=true]{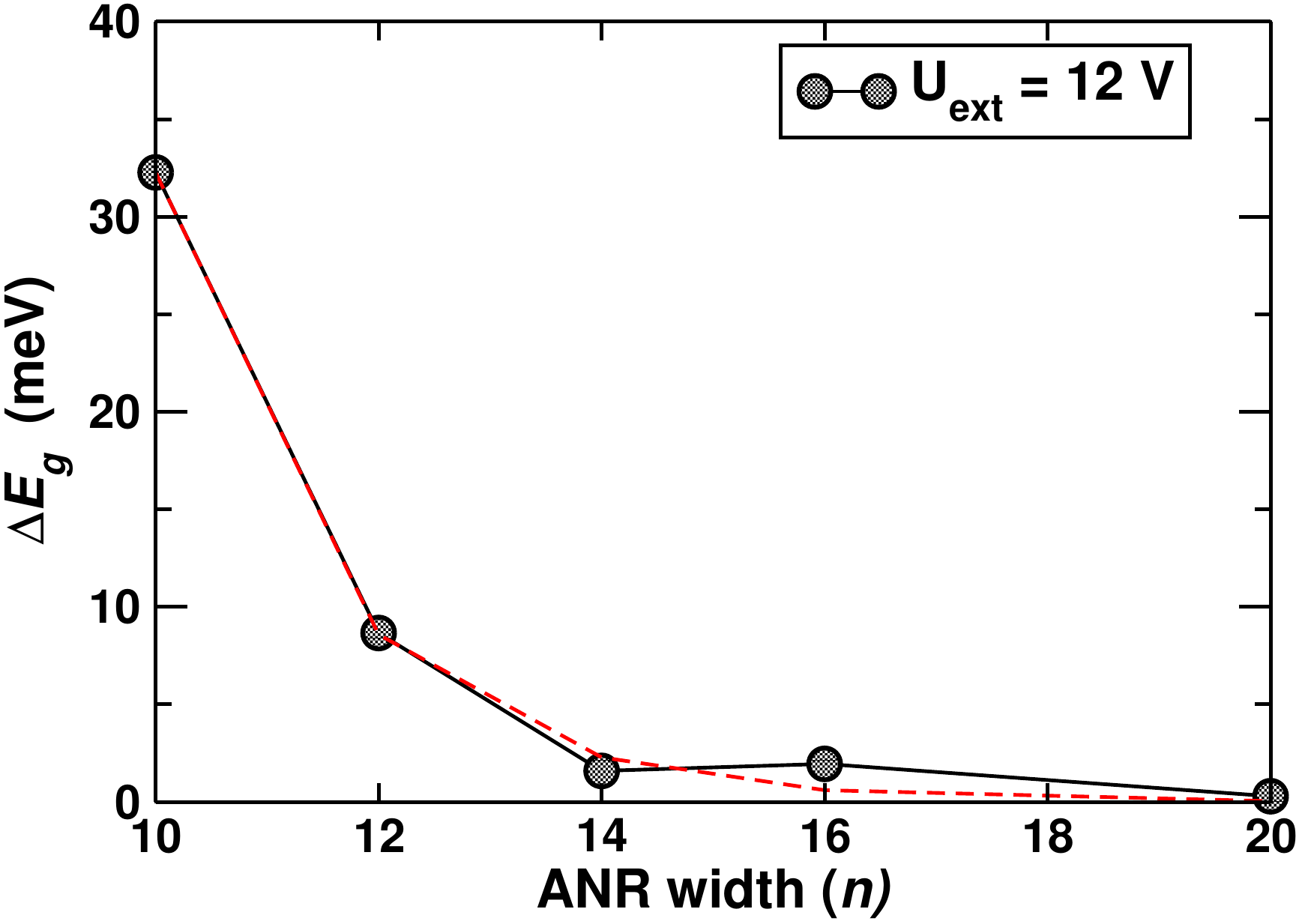}
\caption{Variation of the band-gap $\Delta E_\mathrm{g}$ as a function of the $n$-ANR width, $n$. Results are plotted for an applied external
potential, $U_\mathrm{ext}=12$~V, which is sufficient to shift the band-gap away from $\Gamma$. In this condition $\Delta E_\mathrm{g}$ is a direct
measure of the inter-chain hopping integral $t$. The red dashed line is an exponential fit of the calculated data.}
\label{fig:t_NR}
\end{figure}

Finally, to conclude this section we make a number of addition observations, which further validate our model. Firstly we note that, as 
expected, the interaction between the ribbon's edges gets stronger as the ribbon gets smaller. This means that the band repulsion at the 
band-gap along the $\Gamma$-Y direction strengthens for small nano-ribbons. As a consequence the gap-closure occurs for relatively 
larger fields than those expected by a simple rigid band shift (see figure~\ref{fig:BGvsE_NR}). Secondly, the position in $k$-space of the 
gap immediately before its closure, $k_\mathrm{c}$, moves towards $\Gamma$ as the ribbon becomes wider. This essentially indicates 
that the inter-edge interaction, parameterized with $t$, is reduced for large nano-ribbons. In order to prove such a fact in Fig.~\ref{fig:t_NR} 
we plot the band-gap as a function of the nano-ribbon size $n$. This is calculated for an external electrostatic potential, 
$U_\mathrm{ext}=12$~Volt, sufficient to move the band-gap away from $\Gamma$ for all the ribbons investigated. In this situation the band-gap 
is a direct measure of the inter-edge hopping $t$,  $\Delta E_\mathrm{g}\sim$ 2$t$. Notably the decay is rather severe indicating that 
already for relatively small ribbons ($n>14$) the inter-edge interaction becomes almost negligible.

\begin{figure}
\centering
\includegraphics[width=6.0cm,clip=true]{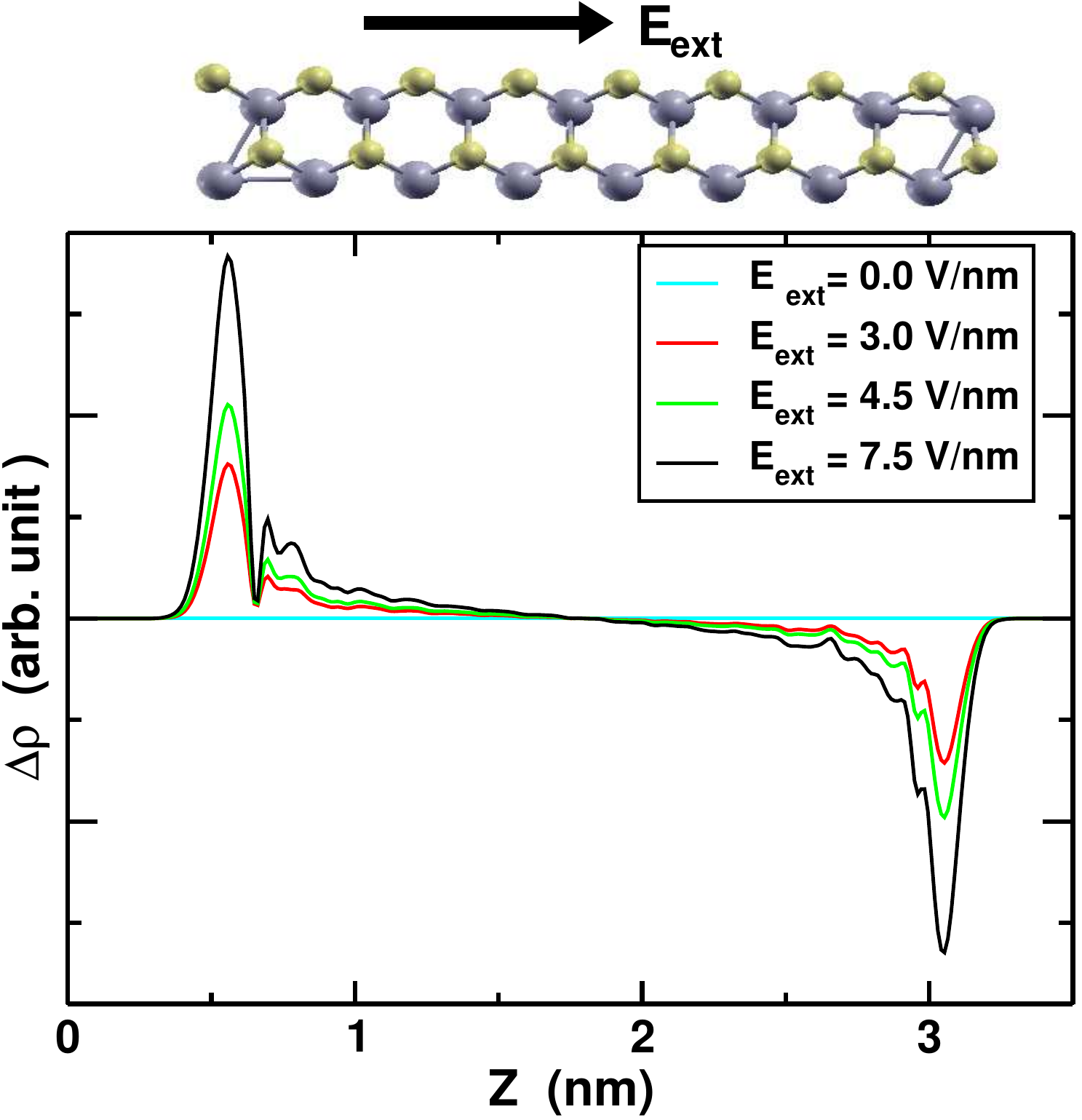}
\caption{(Color online) Field-induced charge density distribution along the nano-ribbon, 
$\Delta\rho={\rho}(E_\mathrm{ext})-{\rho}(E_\mathrm{ext}=0)$, for a 16-ANR and different values of $E_\mathrm{ext}$. $\rho(E_\mathrm{ext})$ is 
the charge density for an external field $E_\mathrm{ext}$ averaged over the longitudinal direction ($xy$-plane) and plotted along the 
transverse one ($z$).}
\label{fig:Rho_NR8}
\end{figure}

Finally we look at the charge density polarization induced in the ANRs by the external electric field. In figure~\ref{fig:Rho_NR8} we plot the 
field-induced charge density distribution, $\Delta\rho$, as a function of the position across the ribbon ($z$-coordinate, see 
Fig.~\ref{fig:10-ANR.eps}). Here $\Delta\rho$ is the difference between the charge density calculated in an applied field $\rho(E_\mathrm{ext})$ 
and that in no field, $\rho(E_\mathrm{ext}=0)$. Also note that all the densities are averaged over the $xy$ plane. Notably there is charge 
accumulation at the positive potential edge and a corresponding depletion at the negative one. As such, a transverse field induces an electrical 
dipole across the ribbon, which effectively behaves as a capacitor. 

\begin{figure}
\centering
\includegraphics[width=6.5cm,clip=true]{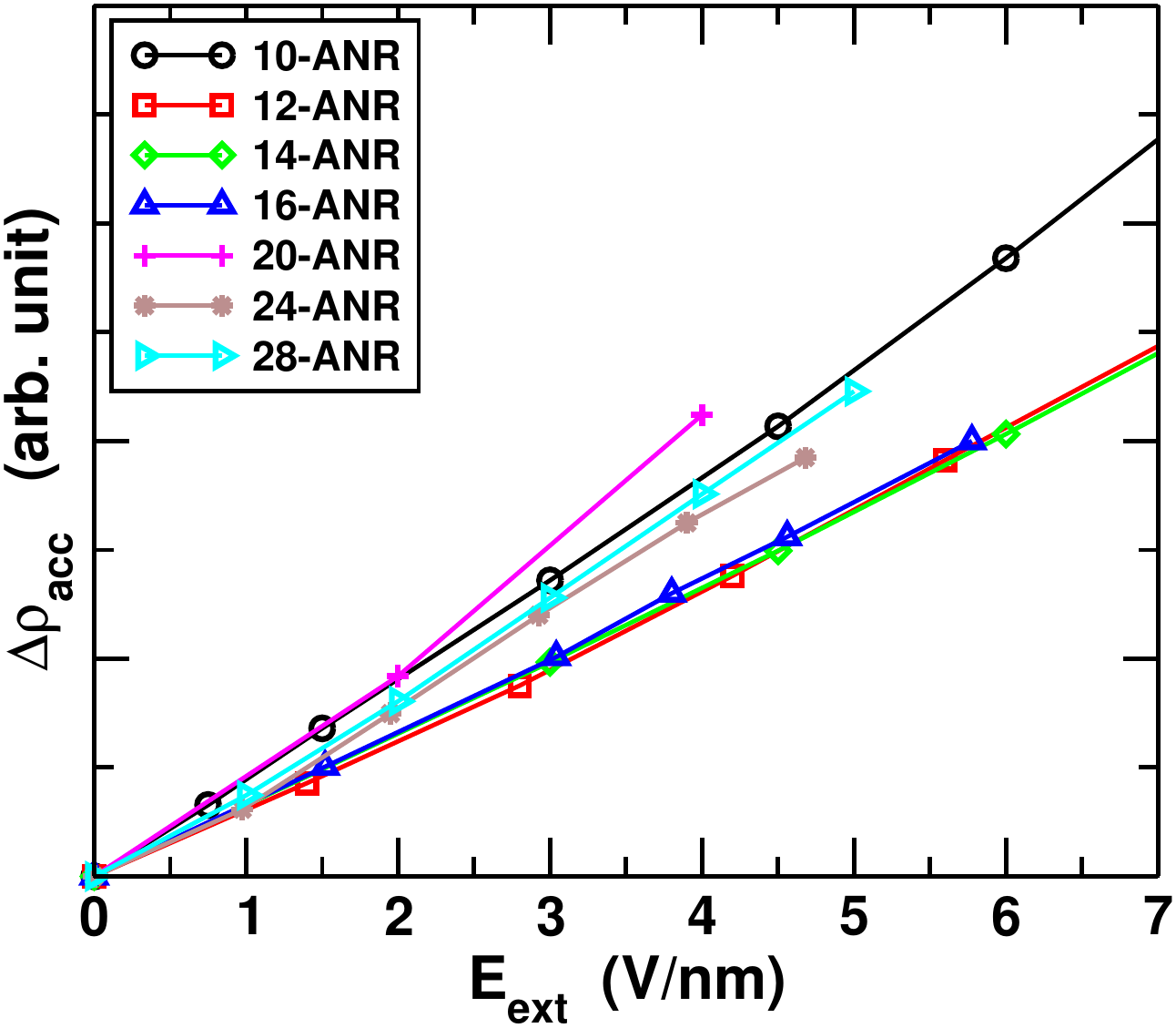}
\caption{(Color online) Charge density accumulation $(\Delta\rho_\mathrm{acc} = \int_{z_\mathrm{L}}^{z_\mathrm{m}} \Delta\rho(z)\:dz$ as 
a function of the external electric field for $n$-ANRs of different width. Note the linear dependence with an almost ribbon-independent slope.}
\label{fig:charge_E}
\end{figure}

The accumulated charge can be calculated by simply integrating $\Delta\rho(z)$ from one of the edge positions, $z_\mathrm{L}$, to the ribbon 
mid-point, $z_\mathrm{m}$, which is $(\Delta\rho_\mathrm{acc} = \int_{z_\mathrm{L}}^{z_\mathrm{m}} \Delta\rho(z)\:dz$. This quantity is presented 
next in Fig.~\ref{fig:charge_E} as a function of $E_\mathrm{ext}$ and for different ANRs. Clearly $\Delta\rho_\mathrm{acc}$ is found to increase 
linearly with the field. This is the behaviour expected from a parallel plate capacitor. A second important observation, also consistent with viewing 
the ribbon as a parallel plate capacitor, is that the slope of the $\Delta\rho_\mathrm{acc}$-$E_\mathrm{ext}$ curve is almost independent from the 
ribbon width. Minor variations can be attributed to our somehow arbitrary definition of the ribbon mid-point (this is defined in terms of the planar 
average of $\Delta\rho(z)$ as the point where $\Delta\rho(z)=0$) and to the fact that as the field increases and the ribbon band-gap
is reduced, the dielectric constant changes. 

In summary the evolution of the electronic properties of MoS$_2$ armchair nano-ribbons as a function of an external transverse electric field 
can be understood in terms of the ribbon dielectric response, which is indeed consistent with that of a linear dielectric. These findings are rather 
general and can be easily transferred to other materials with different band-gaps. Next we examine the effects of stacking multiple MoS$_2$ 
nano-ribbon layers.

\subsection*{Bi-layer and multi-layer MoS$_2$-ANR}\label{Bilayer}
In 2D layered compounds the tiny inter-layer interaction is often sufficient to change drastically the electronic properties of the material. 
A prototypical example is graphene, where the weak $\pi$-$\pi$ interaction is able to turn the linear band-dispersion into parabolic.\cite{BiLayer}
It becomes therefore natural to investigate how the results of the previous section get modified in multi-layered ribbons. 

\begin{figure}
\centering
\includegraphics[width=7.0cm,clip=true]{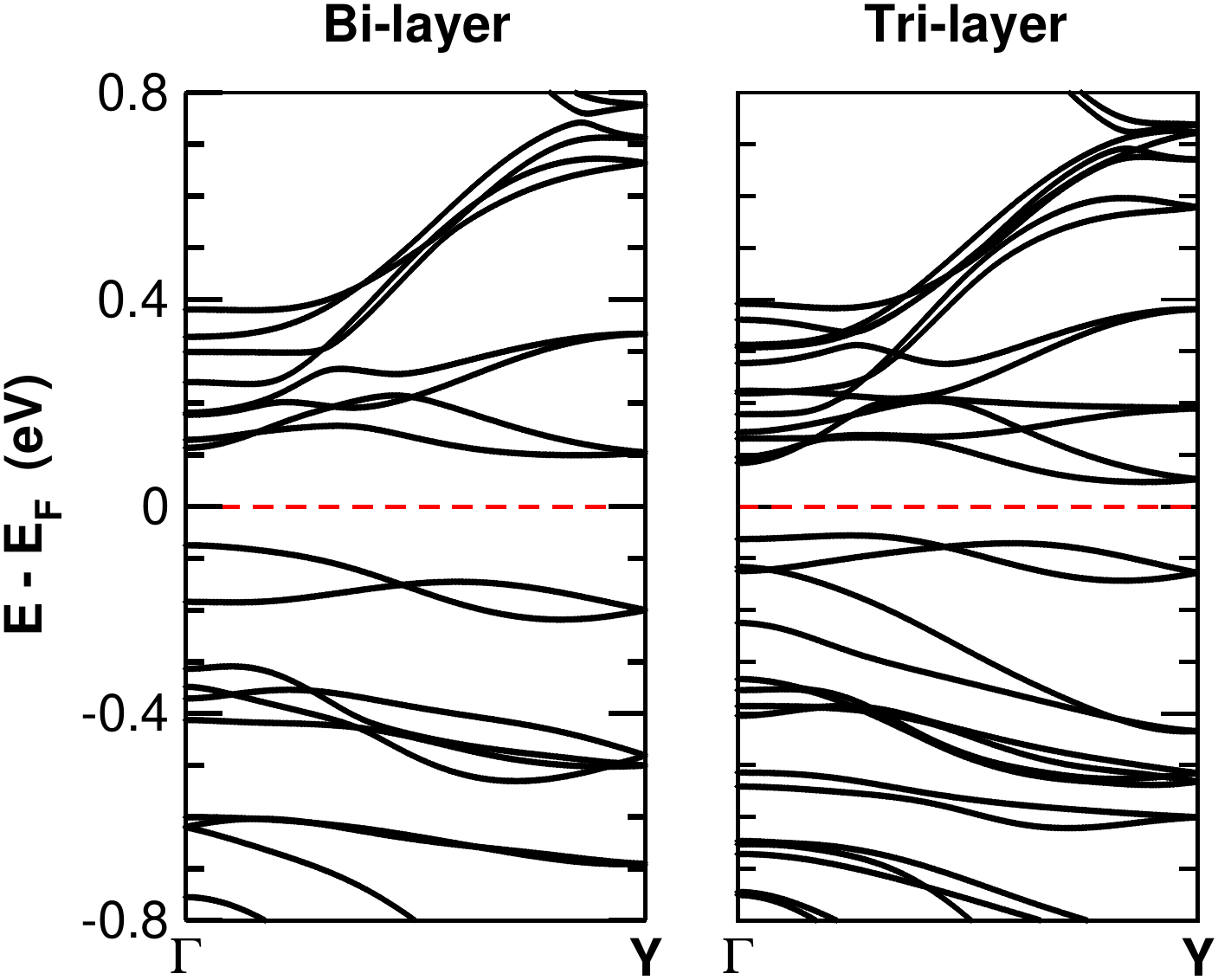}
\caption{The band-structure for a MoS$_2$ 8-ANR bi-layer (left) and tri-layer (right). Note that now the band-gap is indirect with the conduction 
band minimum positioned along the $\Gamma$-Y direction. The horizontal dashed line denotes the position of the Fermi level.}
\label{fig:Band_L2andL3}
\end{figure}

In order to keep the computational costs reasonable we consider here only the case of 8-ANRs, whose electronic band-structure in both a 
bi- and tri-layer form is presented in Fig.~\ref{fig:Band_L2andL3}. As for bulk MoS$_2$ also multi-layered nano-ribbons display an indirect 
band-gap, which is positioned along the $\Gamma$-Y direction in the 1D Brillouin zone. We have extended our calculations to ANRs comprising 
up to five layers and, for comparison, to an infinite (periodic) nano-ribbon stacking. We notice that the band-gap turns indirect already for a bi-layer 
and then it remains indirect for any other structure. Furthermore, the band-gap decreases monotonically with increasing the number of layers. 
However, at variance with their parental 2D counterparts and similarly to the single-layer ribbons, also in multi-layered ANRs both the CB and 
the VB are localized over the ribbon edges. This fact is rather robust with respect to the interlayer-separation, in contrast to what happens to 
the fine details of the electronic structure of the infinite 2D multi-layers,\cite{vdWPRB} so that the explicit inclusion of van der Waals interaction 
in the present context is not crucial. The localization of CB and VB at the edges means that, as in the single-layer case, also the multi-layers 
are sensitive to a transverse electric field.  

\begin{figure}
\centering
\includegraphics[width=7.0cm,clip=true]{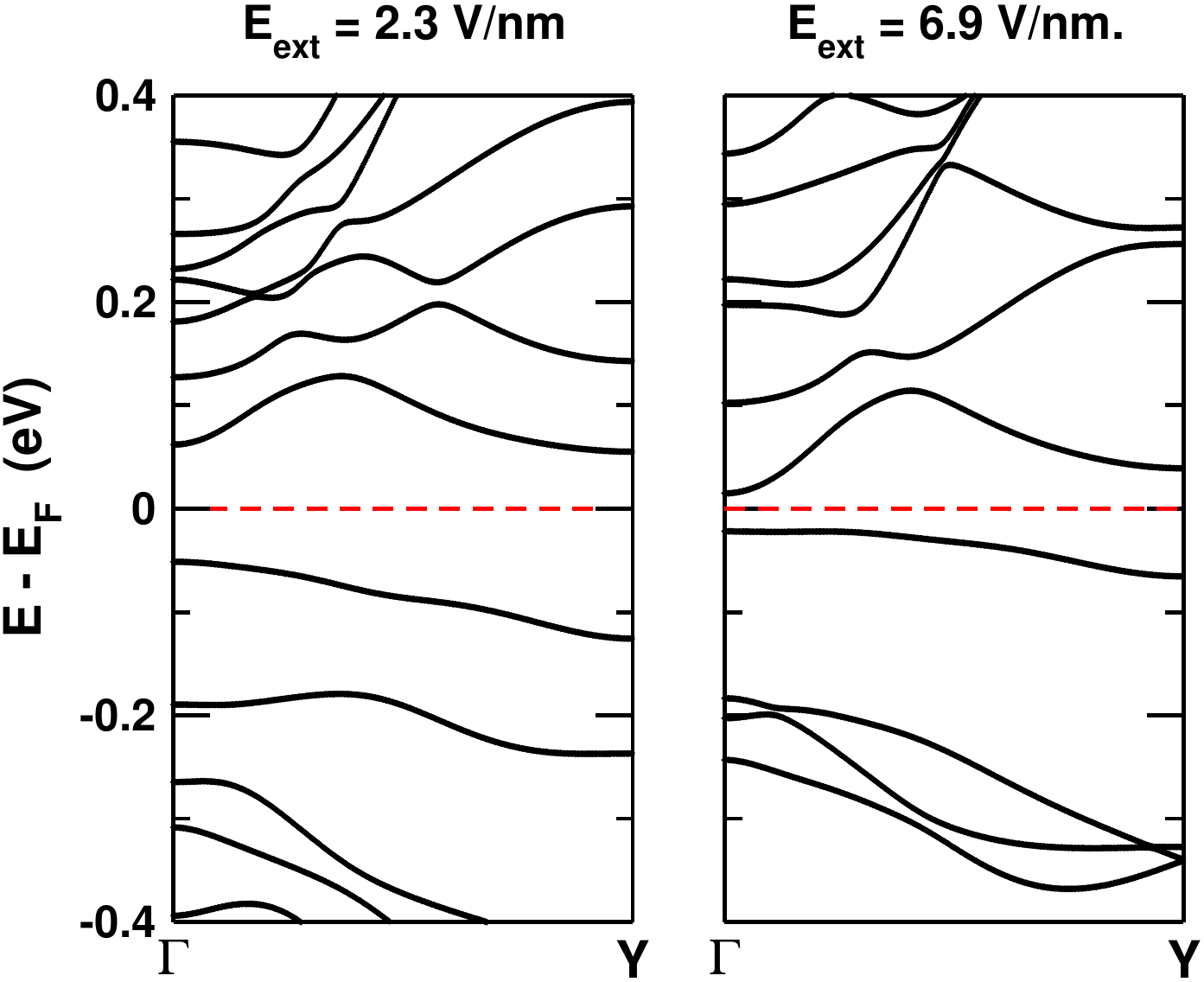}
\caption{The band-structure of a MoS$_2$ 8-ANR bi-layer in the presence of a transverse electric field. For $E_\mathrm{ext}=2.3$~V/nm (left) 
the band-gap is indirect, while it becomes direct at $\Gamma$ for the larger field of $E_\mathrm{ext}=6.9$~V/nm (right).}
\label{fig:Band_L2_2to6}
\end{figure}

Such a sensitivity is examined next for the case of the 8-ANR bi-layer in Fig.~\ref{fig:Band_L2_2to6}. In general the response to 
$E_\mathrm{ext}$ is qualitatively similar to that of the single layers as it is determined by the electrostatic potential shift at the nano-ribbon edge. 
Thus as $E_\mathrm{ext}$ gets larger the energy shift of both the CB and VB localized at opposite edges results in a band-gap reduction. There 
is however a difference with respect to the single layer case, namely that the inter-layer interaction lifts the edge-band-degeneracy for 
$E_\mathrm{ext}=0$. As a consequence the band-dispersion around the band-gap changes in a non-trivial way with the electric field. For 
instance for the case of the bi-layer 8-ANR first the CB minimum moves towards Y, thus strengthening the indirect nature of the gap and then, 
for larger fields, is reverts back to $\Gamma$ and eventually the band-gap becomes direct. This is an intriguing feature as it demonstrates that
in multi-layers the nature of the band-gap can be manipulated by an external transverse field. As such one may expect, for instance, that the optical
activity of such ribbons may be electrically modulated. 

As a final observation we note that the critical fields for closing the gap in multi-layer ANRs are significantly smaller than those needed 
for the corresponding single-layer ones. Furthermore, for a fixed external field the band-gap is found to be inversely proportional to the number 
of layers, although a more precise dependence is difficult to establish. Such an inverse dependence is expected if the different layers in the 
multi-layer ribbon behave effectively like capacitors in parallel, although such an analogy cannot be pushed much further based on our DFT 
results. 

\subsection*{Electrically driven magnetism}\label{MEeffect}
As the VB and the CB of a MoS$_2$-ANR approach each other under the influence of $E_\mathrm{ext}$ a high DOS is generated at the 
Fermi level on the verge of the insulator to metal transition. Such high DOS originates from the Van Hove singularities at the band edges 
owing to the quasi-1D nature of the NRs. For $E_\mathrm{ext}=0$ the ANRs are non-magnetic semiconductors. Thus, for any fields smaller 
than the critical one for the band closure the system remains semi-conducting and no spin-polarized calculations are needed. However, at 
and beyond the onset of the metallic phase both spin polarized and non-spin polarized calculations have been performed in order to establish 
whether the ground state is stable against the formation of a finite magnetic moment. In general, we have found that at the critical electrical 
field, $E_\mathrm{c}$, where the bandgap closes there is a sufficiently high DOS at the (non-magnetic) Fermi level to drive the formation of 
a magnetic moment according to the Stoner criteria for ferromagnetism.
\begin{figure}
\centering
\includegraphics[width=8.0cm,clip=true]{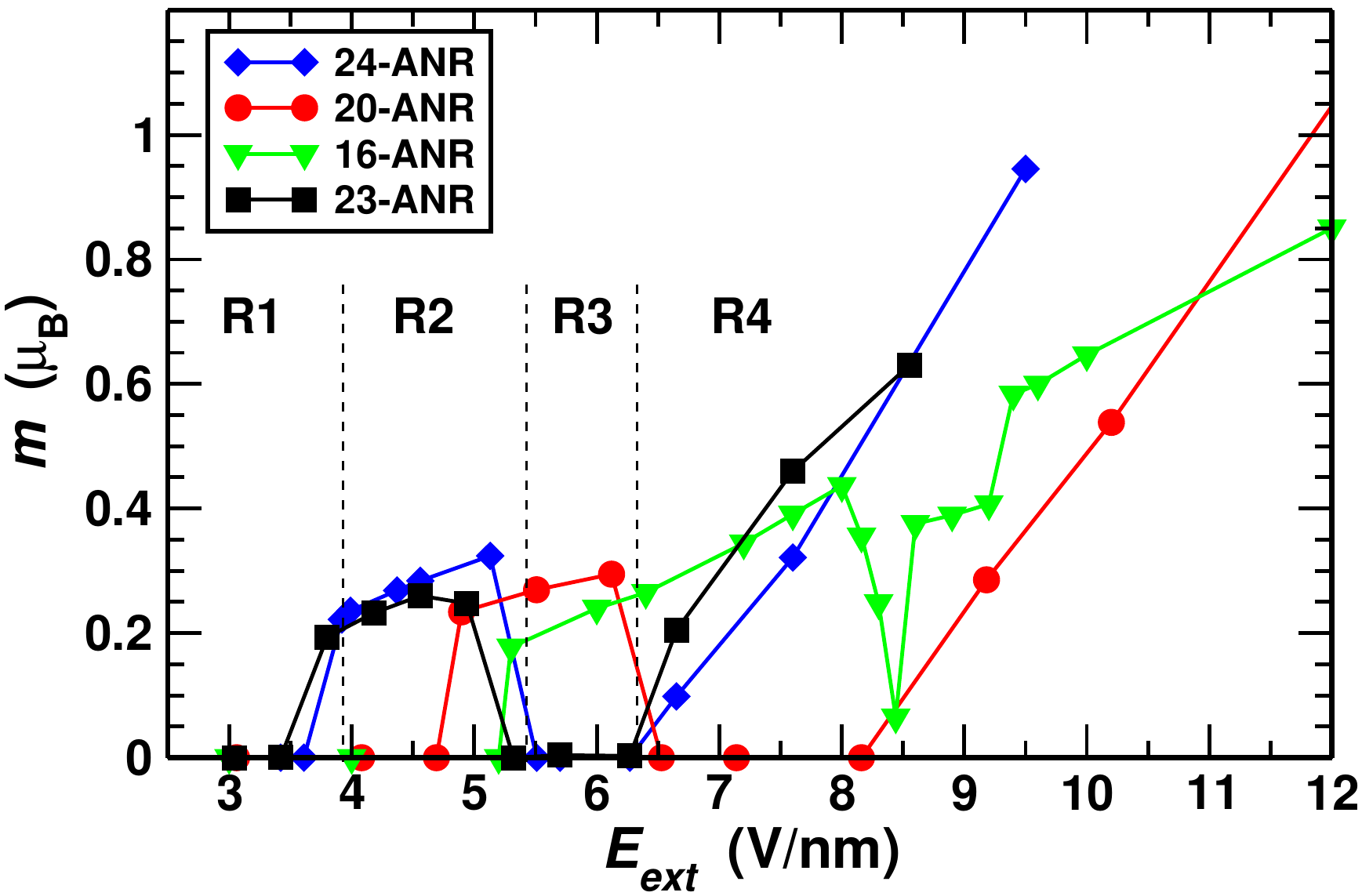}
\caption{(Color online) Variation of the magnetic moment per ribbon cell, $m$, as a function of the external electric field 
$E_\mathrm{ext}$ for 16-ANR (green triangles), 20-ANR (red circles) and 24-ANR (blue diamonds). We also report data for
the symmetric 23-ANR (black squares). R1 and R3 (R2 and R4) define the regions of $E_\mathrm{ext}$ where the 24-ANR 
is in its diamagnetic (magnetic) state.}
\label{fig:spinpolvsE_NRs}
\end{figure}

Figure~ \ref{fig:spinpolvsE_NRs} displays the evolution of the ribbon magnetic moment (per ribbon cell), $m$, as a function of 
$E_\mathrm{ext}$ for three different ANRs. For example, one can clearly see that a 24-ANR becomes magnetic for 
$E_\mathrm{ext}=4$~V/nm, which is the critical field to close the gap completely. Such a transition from a diamagnetic to a magnetic 
ground state is driven by the Stoner criterion, which reads $I\rho_\mathrm{F}>1$, where $I$ is the Stoner parameter 
(the exchange constant) and $\rho_\mathrm{F}$ is the DOS at the Fermi level.\cite{Stoner,Janak} The Stoner parameter can be 
estimated from our DFT calculations, since the magnetic exchange splitting, $\Delta$, of the bands is given by 
$I\bar{m}$,\cite{PRB_1990_41_7028} where $\bar{m}$ is the magnetic moment in units of the Bohr magneton $\mu_\mathrm{B}$. 
For instance, in the case of the 24-ANR at $E_\mathrm{ext}=4.5$~V/nm we find that $\Delta=0.14$~eV and $m=0.27$~$\mu_\mathrm{B}$/unit 
cell, so that the estimated value of the Stoner $I$ parameter is $\sim0.5$~eV and the required DOS at the Fermi level necessary 
to satisfy the Stoner instability condition is $\rho_\mathrm{F}\geq 1/0.5=2$~eV$^{-1}$. 

Also in Fig.~\ref{fig:spinpolvsE_NRs} it can be observed that the critical field for the diamagnetic to magnetic transition corresponds 
exactly to $E_\mathrm{c}$, {\it i.e.} it coincides with the onset of metallicity (see Fig.~\ref{fig:BGvsE_NR}). This is true for both the 20-ANR and 
the 24-ANR and for any larger ribbon. The situation however is different for the 16-ANR for which the magnetic transition occurs at a field 
smaller then $E_\mathrm{c}$. As noted previously, for small ribbons non-spin polarized calculations reveal that the inter-edge interaction 
creates a band anti-crossing, so that the critical field for the band-gap closure is larger than that needed for shifting rigidly the CB and the 
VB by the band-gap. Thus the metallic phase occurs at an $E_\mathrm{c}$ larger than the one necessary to simply shifting the bands. 
In contrast, we find that once spin-polarization is allowed in the calculation, the semiconducting to metallic transition occurs virtually with the 
band edges of the VB and the CB touching at the $\Gamma$ point. This suggests that the exchange energy gained by spin-polarizing the 
system is sufficiently large to overcome the inter-edge interaction. Total energy calculations indeed confirm that the spin polarized ground 
state is energetically more favorable than the diamagnetic one. Hence, the 8-ANR undergoes a magnetic transition at a smaller 
$E_\mathrm{ext}$ that the E$_\mathrm{c}$ calculated from the LDA. 

\begin{figure*}
\centering
\includegraphics[width=16.0cm,clip=true]{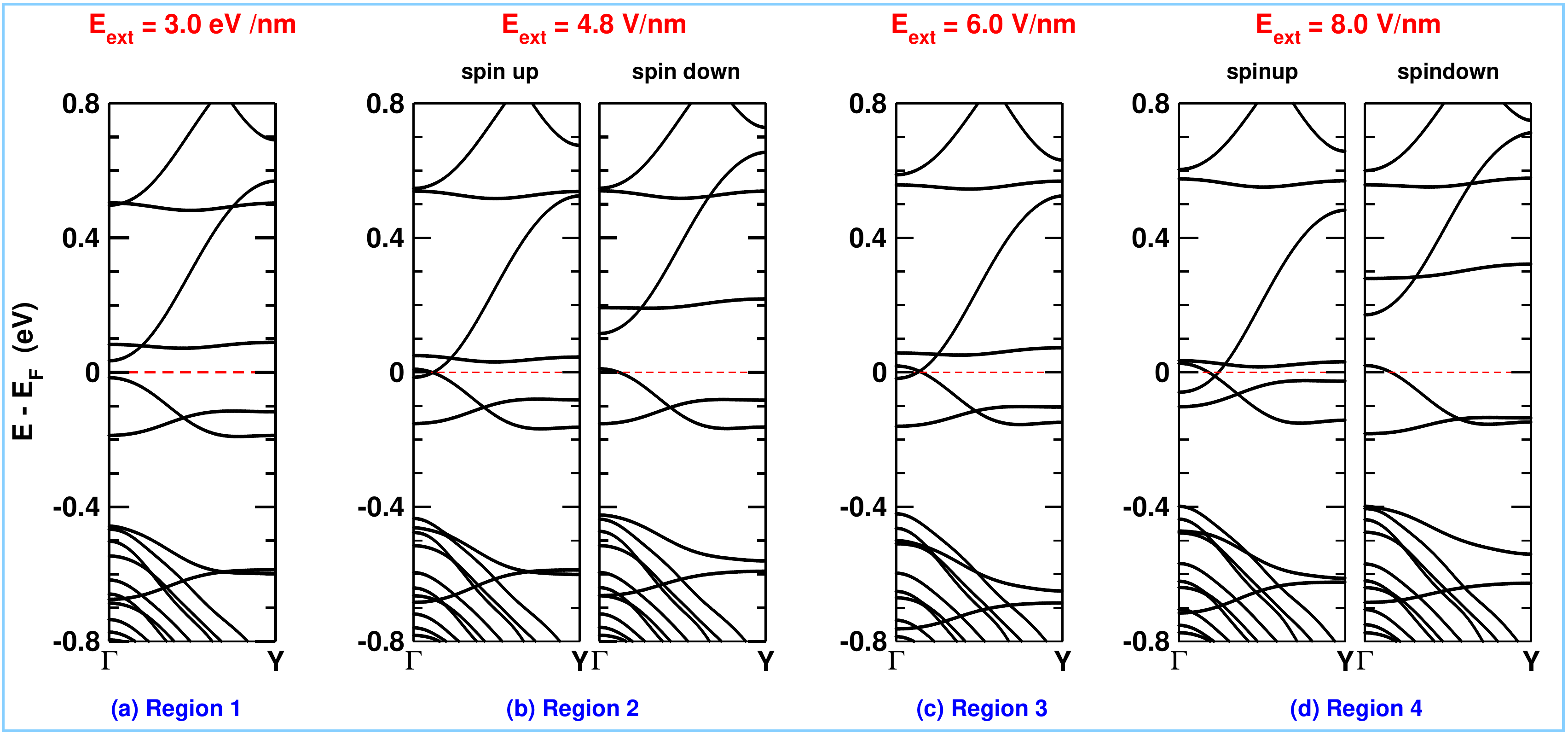}
\caption{(Color online) The band-structure of the 24-ANR plotted for different $E_\mathrm{ext}$. In particular we select four representative 
field strengths corresponding to the four regions defined in Fig.~\ref{fig:spinpolvsE_NRs}.}
\label{band_NR12}
\end{figure*}
It is also interesting to look at what happens when the electric field is increased beyond the value needed for the first magnetic transition. 
Taking the case of the 24-ANR as an example we notice in Fig.~\ref{fig:spinpolvsE_NRs} that there are different magnetic regions depending 
on $E_\mathrm{ext}$. In particular we observe two diamagnetic regions (R1 and R3) and two magnetic ones (R2 and R4). R1 corresponds 
to the semi-conducting ground state of the ribbon and therefore it is non-magnetic. At the boundary between R1 and R2 the ribbon becomes 
metallic and the Stoner mechanism drives the electronic structure in a magnetic state. Further increase of $E_\mathrm{ext}$, however, destroys 
the magnetic moment, which returns to zero in R3. Such a return of the diamagnetic phase can be understood by looking at 
Fig.~\ref{band_NR12}, where we present the band-structure for the 24-ANR at four representative electric field strengths, corresponding 
respectively to the four regions. In R3 the field is strong enough to further shift the CB and VB in such a way that the Van Hove singularities are 
removed from $E_\mathrm{F}$. Now the electronic structure of the ribbon is that of a non-magnetic semi-metal with both an electron and a hole 
pocket at the Fermi level. As $E_\mathrm{F}$ cuts now in a region where the bands have relatively large dispersion (small DOS) the Stoner criterion 
is no longer satisfied and the magnetic moment disappears. A further increase of the external field drives the system into R4, where now a new 
band from the CB manifold crosses the Fermi level and it is spin split by the Stoner exchange. The same mechanism works for
the 20-ANR, while anomalies appear for the 16-ANR, again because of the more subtle inter-edge interaction.

Finally we wish to note that, as previously observed, the fact that the ribbon has inversion symmetry about its axis is irrelevant for the
magnetic moment formation. This is demonstrated again in Fig.~\ref{fig:spinpolvsE_NRs}, where for completeness we report data for the
23-ANR as well. Clearly the 23-ANR and the 24-ANR display an almost identical pattern of magnetic moment formation with $E_\mathrm{ext}$,
except for minor details in the various critical positions for the on-set of the magnetism, which are mainly due to the slightly different
confinement in the two structures. 

\subsection*{Edge termination}
\label{edges}
We finally move to discuss the effects that the different edge terminations have on the onset of the electric field driven magnetism. Experimentally
it was reported that MoS$_2$ single layer clusters present a well defined edge structure.\cite{NatNanoT_2007_2} In particular it was shown 
that clusters above a certain critical size (about 1~nm) all display 100\% rich Mo edges, {\it i.e.} the ones investigated so far throughout our work. 
This broadly agrees with earlier density functional theory calculations,\cite{Schweiger} which however pointed out that alternative edge structures 
can form depending on the clusters growth environment. In particular it was reported that under hydrodesulfurization conditions S-terminated 
edges become more stable. It becomes then meaningful and intriguing to explore whether the results presented so far are robust against 
sulfurization of the edges. 

Towards this goal we have repeated our calculations for the 24-ANR by replacing either one or both of the Mo edges with a different
termination. In particular we have looked at three different cases, namely: 1) single 50\% S-passivated edge, 2) single 100\% S-passivated
edge, 3) double 100\% S-passivated edges. Furthermore for completeness, we have explored whether hydrogen passivation, alternative to 
the S one, produces any qualitative change. 

In general we have found that, regardless of the termination, the valence and the conduction bands of the nanoribbon are always made of
edge states, while their band dispersion and the actual bandgap do depend on the chemical nature of the edges. These two features
suggest that the evolution of the bandgap in an external electric field should present similar qualitative features to those discussed
previously, as the gap closure is simply dominated by the shift in the electrostatic potential at the ribbon edges. In contrast the formation 
of the magnetic moment, which depends on the band dispersion through the density of states and on the exchange interaction of the
edge wave-function, may be sensitively affected by the details of the edge structure.

\begin{figure}[h]
\centering
\includegraphics[width=8.0cm,clip=true]{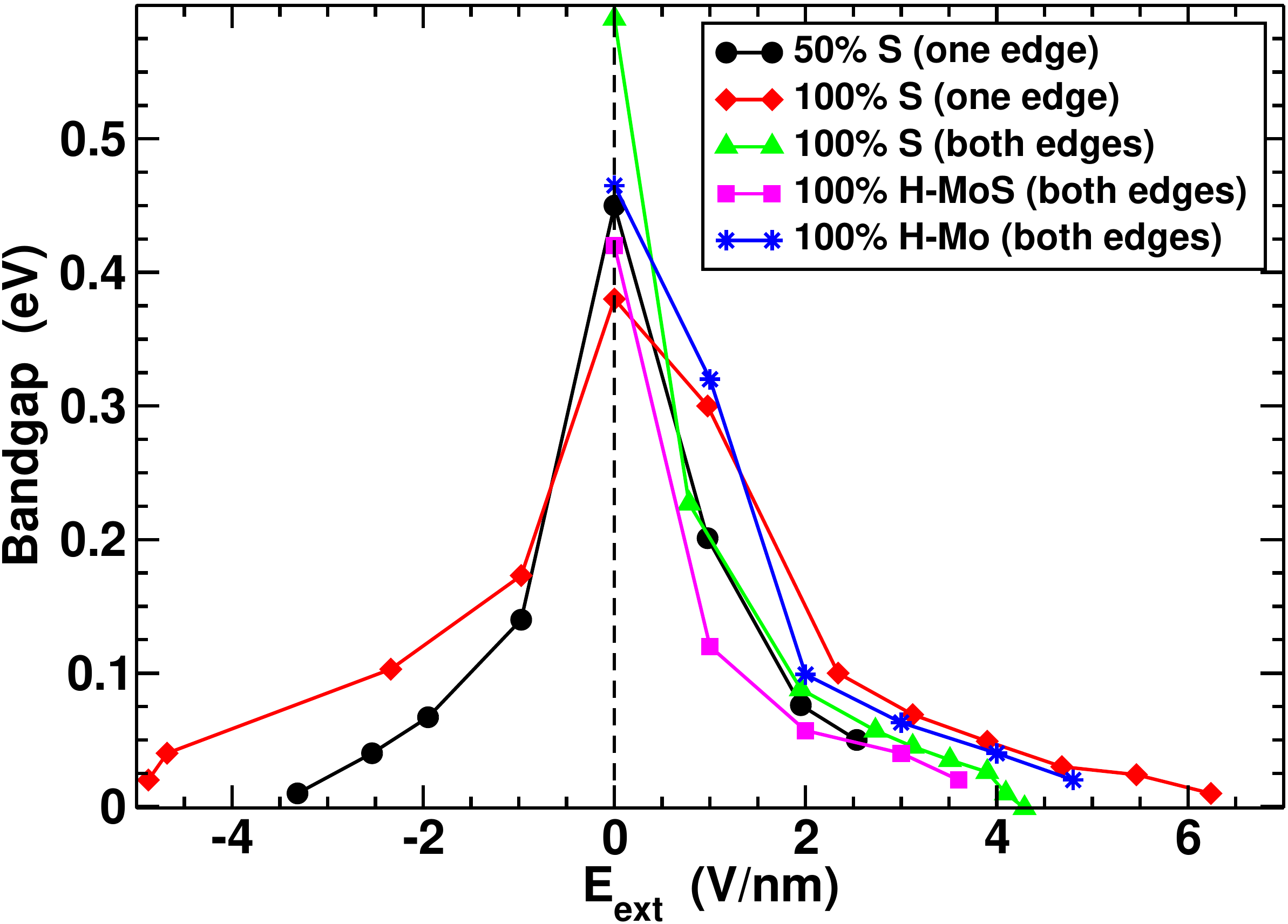}
\caption{(Color online) Evolution of the bandgap in an external electric field of a 24-ANR with different edge terminations. In some cases, 
labelled as ``one edge'', the new termination is only over one of the two edges, while the other remains in the unsaturated configuration 
discussed throughout this work (Mo edge). In this case we plot the gap as a function of field for both the field polarities.}
\label{gap-edges}
\end{figure}
Fig.~\ref{gap-edges} shows the value of the bandgap as a function of the external electric field for all the terminations investigated. Note that,
as some ribbons present different edges, the gap depends not only on the electric field intensity but also on its polarity. As such for these ribbons 
we plot results for both positive and negative $E_\mathrm{ext}$. In general the figure confirms the intuitive picture presented above, {\it i.e.} 
for all the terminations studied we observe gap closure as a function of the electric field. The critical fields are also rather similar ranging
between 4~V/nm (the same critical field for the case of two 100\% Mo rich edges) to approximately 6~V/nm.

\begin{figure}[h]
\centering
\includegraphics[width=8.0cm,clip=true]{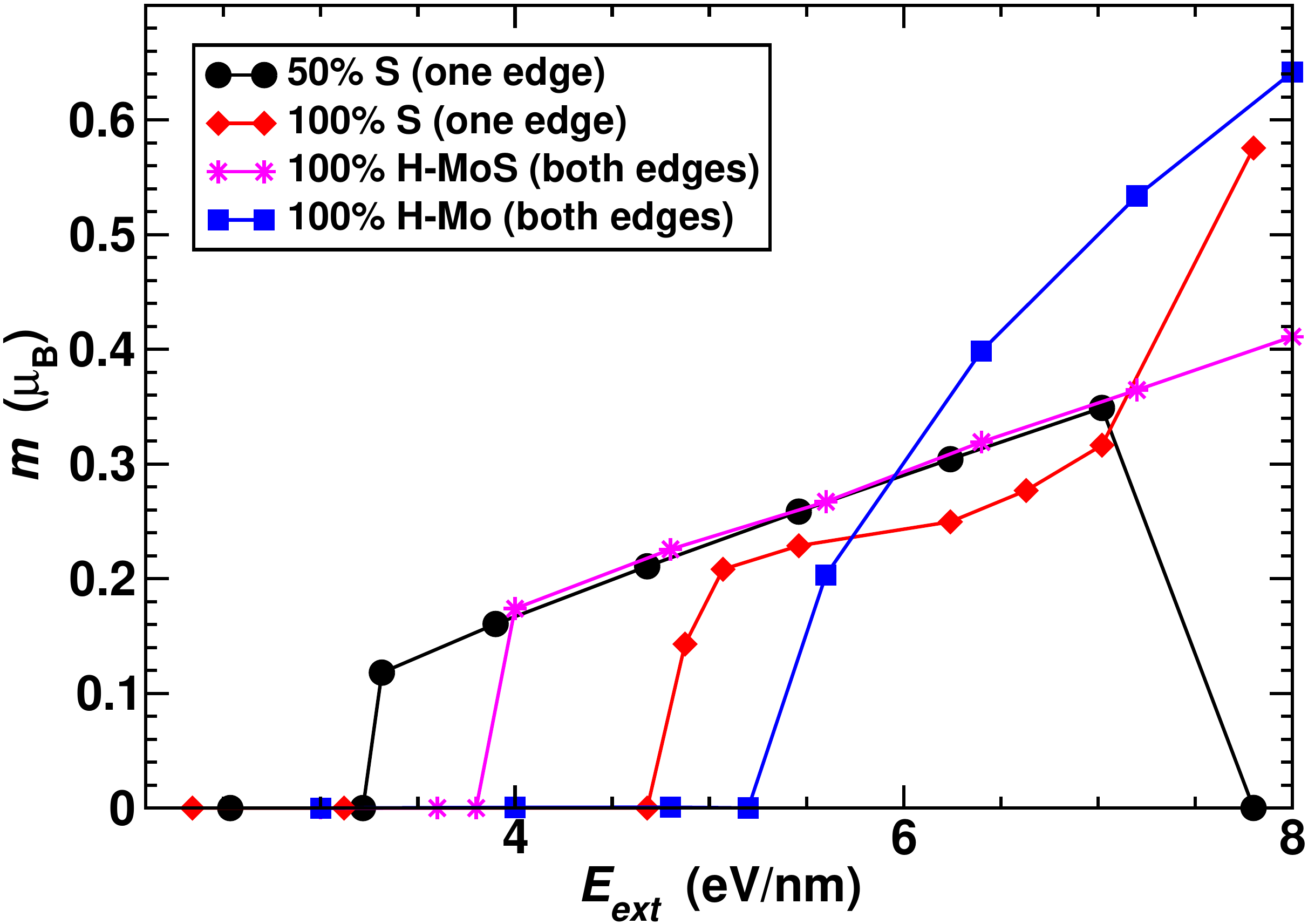}
\caption{(Color online) Magnetic moment as a function of the external electric field for a 24-ANR with different edge terminations. }
\label{Fig15}
\end{figure}

More intriguing is the influence of the edge termination on the formation of the magnetic moment. Here we find that some edges do not
display any Stoner instability so that no magnetism is induced by the external field. This can be appreciated by looking at Fig.~\ref{Fig15},
where we plot the magnetic moment per cell as a function of $E_\mathrm{ext}$.
From the figure it appears that ANRs with either 50\% or 100\% S-rich edges can sustain a magnetic moment. However the figure refers to
a 24-ANR in which only one edge has such termination, while the other still displays the Mo one. A closer look at the density 
of state reveals that the magnetic moment in this case forms only at the Mo edge, but not at the S-rich one. This brings the 
interesting consequence that the moment formation occurs only for one specific polarity of the electric field, which is the one necessary 
to bring the band associated to the Mo edge at the Fermi level. A second consequence is that when the sulfurization is at both 
edges no magnetic moment ever develops. 

We then conclude that in S-rich edges the Stoner condition is not met, either because the density of states is not large enough or because
the additional S contributes to reduce the Stoner parameter of the edge states. Interestingly H passivation does not seem to be detrimental 
to the magnetism. As suggested in Ref.~[\onlinecite{Terrones}] we have investigated two types of passivations differing by whether the
double passivation is only at the Mo site (this is labelled in Fig.~\ref{gap-edges} and Fig.~\ref{Fig15} as ``H-Mo'') or both at the Mo and the 
S ones (labelled as ``H-MoS''). In this case the magnetic moment forms at both edges as soon as $E_\mathrm{exp}$ is large enough to
close the gap, {\it i.e.} H-passivated edges behave identically to the unpassivated Mo ones. 

Finally we conclude with some comments on the possible effects of disorder. Throughout this paper we have investigated only perfect
edges, which is justified given the experimental observation of large clusters with perfect edges.\cite{NatNanoT_2007_2} However these clusters 
count at most approximately 20 sites per side and it is very unlikely that much larger nanoribbons can maintain such structural perfection. 
Defects and inhomogeneities of course break translational invariance so that the one-dimensional nature of the edge states will certainly
be affected. One should then expect a general broadening of the edge-related bands with a consequent reduction of the average density
of states. As such, because the magnitude of the density of states determines the Stoner condition, it is reasonable to expect that the 
formation of the magnetic moment will be rather sensitive to edge defects. In contrast the gap closure in an electric field should be more
robust. This in fact depends only on the ability of creating a potential different between the edges, a feature that should not be affected 
too much by disorder. 

\section*{CONCLUSIONS}
In summary, we have investigated the ground state electronic structure and the electrical field response of MoS$_2$ nano-ribbon structures. 
Our first principle calculations show that MoS$_2$ ANRs are insulators with a direct band-gap regardless of the width. Importantly the band-gap 
in these systems is primarily determined by a pair of edge states and it may be tuned by applying an external transverse electric field. This 
can eventually drive a metal-insulator transition. It is important to note that the critical electric field for the transition can be reduced to a practical 
range with increasing ribbon width. Also it is interesting to remark that, as the dielectric constant is approximately proportional to the inverse of 
the band-gap, the critical fields for the gap closure are expected to be relatively materials independent. 

The presence of localized edge states that can be moved to the Fermi level suggests that the system can be driven towards magnetic instability. 
Our spin-polarized calculations show that this indeed happens and that at a certain critical electric field a diamagnetic to magnetic transition 
occurs. This follows directly from the Stoner criterion as the Van Hove singularities associated to the edge states have a large density of 
states. Intriguingly the magnetic phase can be further tuned by the external field and different alternating diamagnetic and magnetic regions 
can be accessed. \\

\section*{METHODS}
Electronic structure calculations are performed by using density functional theory (DFT) \cite{PRB_1964_136, PRA_1965_140} and the 
Ceperly-Alder 
parametrization~\cite{PRL_1980_45} of the local spin density approximation (LSDA) to the exchange and correlation functional. In particular 
we employ the {\sc Siesta} code.\cite{IOP_2002_14} A double-$\zeta$ polarized~\cite{Siesta_Basis} numerical atomic orbital basis set for Mo 
and S is used together with the Troullier-Martins scheme for constructing norm-conserving pseudopotentials.\cite{PRB_1991_43} The
pseudopotentials are generated by treating the following electronic states as valence: Mo: $5s^1 5p^0 4d^5 4f^0$; S: $3s^2 3p^4 3d^0$. An 
equivalent plane wave cutoff of 250~Ry is chosen for the real space grid and the Brillouin zone is sampled by using a $(1\times100\times1)$ 
Monkhorst-Pack grid. Periodic boundary conditions have been included and a vacuum layer of at least 15~{\AA} is placed at the edges of the 
ribbon both in plane and out of plane in order to suppress the interaction between the ribbon periodic images. Conjugate gradient is used to 
obtain optimized geometries, where all the atoms in the unit cell are allowed to relax under the action of the external electric field until the forces 
on each atom are less than 0.03~eV/\AA.

\section*{Acknowledgments} This work is supported by Science Foundation of Ireland (Grant No. 07/IN.1/I945). We thank Irish Centre for High 
End Computing (ICHEC) and Trinity Centre for High Performance Computing (TCHPC) for the computational resources provided.

\end{document}